\documentclass[%
 reprint,
 amsmath,amssymb,
 superscriptaddress,
 aps
]{revtex4-2}

\usepackage{graphicx} 
\usepackage{dcolumn} 
\usepackage{bm} 
\usepackage{enumerate}
\usepackage{mathtools}
\usepackage{physics}
\usepackage{booktabs}
\usepackage{soul}
\usepackage{enumitem}
\usepackage{subcaption}
\usepackage{aas_macros}

\makeatletter
\let\MYcaption\@makecaption
\makeatother
\makeatletter
\let\@makecaption\MYcaption
\makeatother
\usepackage{hyperref}
\usepackage{xcolor}

\newif\ifDraft
\Draftfalse

\definecolor{DarkGreen}{rgb}{0.0,0.45,0}
\definecolor{Sakujo}{rgb}{0.9,0.5,0.9}
\definecolor{Grey}{rgb}{0.7,0.7,0.7}
\ifDraft

\def\red#1{\textcolor{red}{#1}}

\else

\def\red#1{#1}

\fi

\begin{document}


\title{Gravitational wave memory from accelerating relativistic jets \\ in multiple thick shell scenarios} 

\author{Yusuke Sakai}
\affiliation{Department of Design and Data Science and Research Center for Space Science, Advanced Research Laboratories, \\ Tokyo City University, 3-3-1 Ushikubo-Nishi, Tsuzuki-ku, Yokohama, Kanagawa 224-8551, Japan}%

\author{Ryo Yamazaki}
\affiliation{Department of Physical Sciences, Aoyama Gakuin University, 5-10-1 Fuchinobe, Sagamihara 252-5258, Japan}%
\affiliation{Institute of Laser Engineering, Osaka University, 2-6 Yamadaoka, Suita, Osaka 565-0871, Japan}%

\author{Yoshihiro Okutani}
\affiliation{Department of Physical Sciences, Aoyama Gakuin University, 5-10-1 Fuchinobe, Sagamihara 252-5258, Japan}

\author{Satsuki Ueno}
\affiliation{Department of Physical Sciences, Aoyama Gakuin University, 5-10-1 Fuchinobe, Sagamihara 252-5258, Japan}

\author{Norichika Sago}
\affiliation{Division of General Education, Kanazawa Medical University,
1-1 Daigaku, Uchinada, Kahoku, Ishikawa 920-0293, Japan}%

\author{Marco Meyer-Conde}
\affiliation{Department of Design and Data Science and Research Center for Space Science, Advanced Research Laboratories, \\ Tokyo City University, 3-3-1 Ushikubo-Nishi, Tsuzuki-ku, Yokohama, Kanagawa 224-8551, Japan}%
\affiliation{University of Illinois at Urbana-Champaign, Department of Physics, Urbana, Illinois 61801-3080, USA}

\author{Hirotaka Takahashi}
\affiliation{Department of Design and Data Science and Research Center for Space Science, Advanced Research Laboratories, \\ Tokyo City University, 3-3-1 Ushikubo-Nishi, Tsuzuki-ku, Yokohama, Kanagawa 224-8551, Japan}%
\affiliation{Institute for Cosmic Ray Research (ICRR), The University of Tokyo, 5-1-5 Kashiwa-no-Ha, Kashiwa City, Chiba 277-8582, Japan}
\affiliation{Earthquake Research Institute, The University of Tokyo, 1-1-1 Yayoi, Bunkyo-ku, Tokyo 113-0032, Japan}

\date{\today}

\begin{abstract}

Gravitational wave (GW) memory, a permanent distortion of the space-time metric, is anticipated during the acceleration of relativistic jets in gamma-ray bursts (GRBs).
While the precise mechanism behind GRBs is not yet fully understood, detecting GW memory may contribute to clarifying their nature.
In this paper, we consider various scenarios of GW memory emission, including both single and multiple shells with thin and thickshells. 
In particular, the memory spectrum for each scenario is compared with the sensitivity of next-generation detectors, namely Deci-hertz Interferometer Gravitational-Wave Observatory and Einstein Telescope.
Physical properties spread over a broad-band region, emphasizing the importance of combined and wide-band observations.
We also simulate GW memory based on 
nearby, realistic scenarios and demonstrate its detectability.

\end{abstract}


\maketitle

\section{Introduction\label{sec:intro}}

Gamma-ray bursts (GRBs) are high-energy astrophysical phenomena in which intense gamma rays come from the source at cosmological distances in a short time period (see Ref.~\cite{ref:zhang2018} for review). 
Gamma rays are emitted by narrowly collimated relativistic jets. 
However, the details of the jet launch mechanism are not fully understood because electromagnetic waves emitted from the jet acceleration region near the central engine are screened by 
dense matter, preventing direct observation.
Observations of gravitational waves (GWs) help us to understand 
the physical processes near the central engine.

For a GRB jet with the bulk Lorentz factor greater than $\sim10^2$, the jet acceleration involving a large mass may emit detectable GW memories \cite{ref:segalis2001,ref:piran2002,ref:sago2004,
ref:Birnholtz2013,ref:piran2021,ref:piran2022arXiv}.
In such cases, the metric perturbation due to the GW permanently deviates from its original value and remains. This imprint is known as the GW memory effect.
The GW memory provides information about the jet structure, acceleration processes, 
and the history of the mass ejection from the central engine.
Relativistic jets may also arise in core-collapse supernovae
\cite{ref:Piran2019ApJ},
binary neutron star mergers \cite{ref:Mooley2018},
tidal disruption events \cite{ref:Bloom2011},
and even in giant flares of magnetars 
\red{
\cite{ref:Yamazaki2005,ref:piran2022arXiv}.
}
Some of these jets are choked during propagation through the dense surrounding medium, and in such a case, a bright electromagnetic signal is unlikely.
Therefore, the detection of such GW memory induced by these astrophysical phenomena will give us not only the evidence for the relativistic jets but also the mechanism of such high-energy astrophysical phenomena.
Note that in this paper, we only consider baryonic matter-dominated jets accelerating to ultrarelativistic speeds, while the GW memory effect may also arise from nonbaryonic components, such as neutrino-driven jets \cite{ref:Suwa2009} and also 
nonrelativistic aspherical supernovae 
\cite{ref:Kotake2006,ref:Murphy2009,ref:Richardson2024}.

Theoretical studies on GW memory emission during the initial jet acceleration phase have progressed as follows.
\citeauthor{ref:segalis2001}~\cite{ref:segalis2001}
considered GW memory induced by a point particle that was 
ejected from the central engine and
accelerated to ultrarelativistic speed. \citeauthor{ref:sago2004}~\cite{ref:sago2004} extended a formalism for an infinitesimally thin shell with finite angular size, expressing the collimated jet.
They simulated emission from multiple shells using step functions as the expression of GW memory, meaning each shell accelerates instantaneously. 
\citeauthor{ref:Birnholtz2013}~\cite{ref:Birnholtz2013} relaxed the assumption of instantaneous acceleration, and calculated the signal from the continuously accelerating axisymmetric jet.
Furthermore, \citeauthor{ref:piran2021}~\cite{ref:piran2021} considered  more general cases involving prolonged acceleration and prolonged jet ejection from the central engine.

After the initial acceleration, the jet propagates into the dense surroundings such as the progenitor star in long GRBs or merger ejecta in short GRBs,
and changes its velocity, resulting in an additional shift of the metric perturbations.
Using thin shell approximation,
\citeauthor{ref:Yu2020}~\cite{ref:Yu2020} calculated GW memory from the jet interacting with the star's envelope or the ejecta.
Moreover, \citeauthor{ref:Urrutia2023}~\cite{ref:Urrutia2023} performed hydrodynamic simulations of jets that penetrate collapsing stars. In their simulation, the preaccelerated jet is injected at the inner boundary at $5\times10^8$~cm from the source, with an initial Lorentz factor of 10. The time evolution of the GW memory is derived, highlighting a double peak in the time domain; one originates from the initial jet acceleration to the final Lorentz factor, and the other is from the jet-stellar envelope interaction near the low-density surface envelope of the star.
Later, the jet breaks out the dense material and 
begins to emit gamma rays. 
The dissipation of the jet's kinetic energy into photons
occurs far from the central engine ($\gtrsim10^{13}$~cm), which again causes the variation of the metric perturbation.
Taking into account the photon emission,
\citeauthor{ref:Akiba2013}~\cite{ref:Akiba2013}
calculated the GW memory from the multiple massive points colliding with each other.
They were interested in the simultaneous detection of gamma-ray photons and GW memory so that the jet was viewed on axis; that is, the angle between the jet axis and the observer's line of sight was less than the inverse of the bulk Lorentz factor of the jet. 
In such a case, the antibeaming effect 
\cite{ref:Birnholtz2013,ref:sago2004} reduces the observed GW amplitude.
Note that detecting the GW memory requires the source to be nearby, which increases the likelihood of 
viewing the jet off axis.
However, the GW memory may still be detectable due to the antibeaming effect.
For nearby off axis viewing events,
observed photon flux is 
still detectable despite of the relativistic beaming effect~\cite{ref:Ioka2001,ref:Yamazaki2002,ref:Yamazaki2003,
ref:Yamazaki2004,ref:Donaghy2006,ref:Salafia2015}.
Moreover, the GW amplitude can be larger for the off axis cases
than the on axis cases.
Finally, in \citeauthor{ref:Huang2023}~\cite{ref:Huang2023}
studied the GW memory from a thin-shell jet with late-time energy injection to account for the observed shallow decay phase in the canonical X-ray afterglow \cite{ref:Nousek2006}.

Gamma-ray light curves for long GRB consist of many pulses (see Ref.~\cite{ref:Piran1999PhR} for a review).
According to the internal shock model of the prompt gamma-ray emission \cite{ref:Rees1994,ref:Sari1997}, the density in the jet is highly inhomogeneous in the radial direction, and the high-density regions, which we refer to as shells hereafter, have different velocities, so that the shells collide with each other, causing shocks that accelerate gamma-ray emitting high-energy electrons.
A collision of two shells produces a pulse in gamma-ray light curves. Since typical long GRBs have around $10^2$ pulses, the jets are thought to contain approximately $10^2$ shells.
The finite rising timescale of the gamma-ray pulse reflects the lifetime of the internal shocks propagating into the shell with finite thickness.
Taking into account the internal shock model, 
\citeauthor{ref:Akiba2013}~\cite{ref:Akiba2013}
calculated GW memory in the point mass approximation.
In this paper, extending previous theoretical works
\cite{ref:sago2004,ref:Birnholtz2013,ref:piran2021,ref:Akiba2013}, we consider the GW memory from an accelerating jet that consists of 
many thick shells.
In \citeauthor{ref:sago2004}~\cite{ref:sago2004},
the jet consists of many patchy, infinitesimally thin shells accelerating instantaneously.
\citeauthor{ref:Birnholtz2013}~\cite{ref:Birnholtz2013} considered GW memory emission from an angularly structured thin shell.
In \citeauthor{ref:piran2021}~\cite{ref:piran2021},
various cases are considered, one of which is a jet 
with a thick shell formed by continuous mass ejection.
The present work builds upon the study conducted in~\cite{ref:piran2021}, with a particular emphasis on the role of radial inhomogeneity. We calculate the observed spectrum of GW memory in specific cases and discuss its detectability by next-generation detectors, such as Deci-hertz Interferometer Gravitational-Wave Observatory (DECIGO)~\cite{ref:Kawamura2008} and the Einstein Telescope~(ET-D)~\cite{ref:Punturo2010}.

The paper is organized as follows. 
In Sec.~\ref{sec:gw-memory}, we introduce our emission 
model of GW memory from GRB jets. The waveform implementation of GW memories under various conditions is presented in Sec.~\ref{sec:experiment}. GW memories are simulated under nearly realistic conditions in Sec.~\ref{sec:practical}. Finally, we summarize our study and discuss its astrophysical aspect in Sec.~\ref{sec:summary}. The overview of next-generation detectors is introduced in Appendix~\ref{app:next-gw-detectors}. The derivation of several formulas is explained in Appendix~\ref{app:gwm}.


\section{Waveform formulation}\label{sec:gw-memory}

\begin{figure}[t]
\includegraphics[width=\columnwidth]{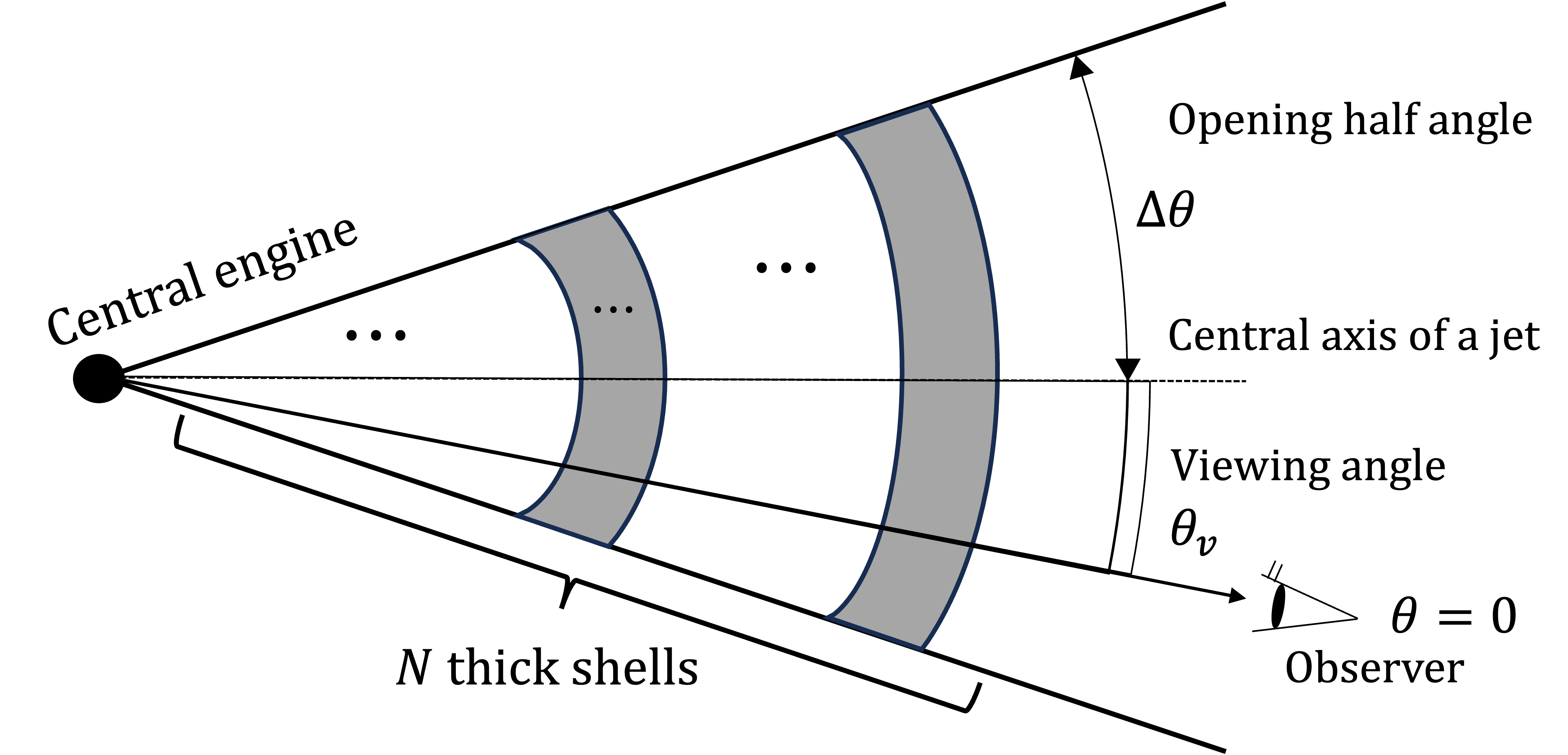}
\caption{An illustration of an accelerating jet.
The observer is in the direction $\theta=0$.
The jet consists of $N$ thick shells.
}
    \label{fig:grb-geometry}
\end{figure}

According to the internal shock model of the prompt GRB emission, we consider an accelerating jet with $N$ thick shells. 
We take the spherical coordinate system $(t,r,\theta,\phi)$ in the rest frame of the central engine. As shown in Fig.~\ref{fig:grb-geometry}, each shell shares a common opening half angle $\Delta\theta$ and the viewing angle between the observer and the jet central axis $\theta_v$, where $\theta = 0$ means the direction toward the observer. 
The shells emerge at $r=r_0$ and subsequently begin to accelerate.
In the following, we measure the Lorentz factor of the moving shells, $\gamma=1/\sqrt{1-\beta^2}$, where $\beta$ is their velocity divided by the speed of light $c$, in the central engine frame.

We approximate 
the GW memory
emission of a thick shell as the superposition of emission $n$ infinitesimally thin shells, with sufficiently large $n$ (finally, we apply $n \to \infty$),
Hence, before addressing the case of thick shells, we will elaborate on the multiple thin shell case.


\subsection{Single / multiple thin shells}
Extending previous studies \cite{ref:segalis2001,ref:sago2004,ref:piran2021},
we calculate the GW memory from accelerating shells in finite time.
A jet consists of $N$ thin shells, and the $j$-th shell ($j=1,2,\cdots, N$) has a terminal Lorentz factor $\gamma_j$ and the total mass $m_j$. 

In order to capture the typical timescale of the shell acceleration, let us consider a simple case of constant acceleration
\cite{ref:Goodman1986ApJ,ref:Paczynski1986,ref:shemi1990appearance},
$\gamma(r)\propto r$.
Suppose the acceleration of the $j$-th shell occurs between
$r=r_0$ and $r=\gamma_jr_0$.
Then, the acceleration time is given by:
\begin{equation}
    t_\mathrm{acc}(\gamma_j) = \frac{r_0}{c} \sqrt{\gamma^2_j - 1},
    \label{eq:tacc}
\end{equation}
in the central engine frame (see Appendix~\ref{app:gwm}-\ref{eq:tacc}).
The observed acceleration time, $T_\mathrm{obs}(\theta,\gamma_j)$, 
for the mass element
moving in the direction $\theta$ from the jet central axis is different 
from $t_\mathrm{acc}(\gamma_j)$ because of the propagation effect, 
and it is given by (see Appendix~\ref{app:gwm}-\ref{eq:tobs})
\begin{equation}
    T_\mathrm{obs}(\theta, \gamma_j) = \frac{r_0}{c}\left(
    \sqrt{\gamma^2_j - 1} - (\gamma_j - 1) \cos\theta
    \right).
    \label{eq:tobs}
\end{equation}
Hence, we get an order of magnitude for $T_\mathrm{obs}(\theta, \gamma_j)$:
\begin{equation}
    \begin{split}
    T_\mathrm{obs}(\theta, \gamma_j) &\sim \frac{r_0}{c} \left(
    1 - \frac{1}{2\gamma_j}+\frac{\gamma_j\theta^2}{2}
    \right)
    \sim \mathcal{O}\left(\frac{r_0}{c}\right),
    \end{split}
    \label{eq:order-tobs}
\end{equation}
in the case of the ultrarelativistic limit ($\gamma_j\gg1$) 
and small $\theta$.

The observed GW memory from a point mass in the $j$-th shell that is accelerated in the direction $(\theta,\phi)$ over a finite time interval, $T_{\rm obs}(\theta,\gamma_j)$, is supposed to be 
\begin{widetext}
\begin{equation}
    h_{\mathrm{point}}(T,\theta,\phi,\gamma_j)  = \frac{2 \gamma_j \beta^2(\gamma_j) m_jG}{c^2 R \Delta \Omega}  \frac{\sin^2 \theta \cos2\phi}{1 - \beta \cos \theta }\eta(T ,\theta,\gamma_j),
    \label{eq:step-gw}
\end{equation}
where $\beta(\gamma)=\sqrt{1-1/\gamma^2}$ hereafter, and $\Delta \Omega = 2\pi (1 - \cos\Delta\theta)$ is the solid angle 
of the shell. In this context, a point mass is calculated as $m_j/\Delta\Omega$.

We simply adopt a linear function for the rising shape of the observed GW memory:
\begin{equation}
\eta(T ,\theta,\gamma_j) =
\begin{cases}
    \ 0 & (T<T_s(\theta)), \\
    \ (T-T_s(\theta))/ {T_\mathrm{obs}(\theta,\gamma_j)} & (T_s(\theta)<T<T_\mathrm{E}(\theta,\gamma_j)), \\
    \ 1  & (T>T_\mathrm{E}(\theta,\gamma_j))~~.
\end{cases}
\end{equation}
\end{widetext}
Here, the metric perturbation at the observer starts to change at
$T=T_s(\theta)=r_0(1-\cos\theta)/c$ and ends at $T=T_E(\theta,\gamma_j)=T_s(\theta)+T_{\mathrm{obs}}(\theta,\gamma_j)$,
and hereafter we adopt Eq.~(\ref{eq:tobs}) for the functional form
of $T_{\mathrm{obs}}(\theta,\gamma_j)$.
Time zero of the observer time, $T$, is determined as follows. Consider a ``hypothetical'' point particle moving toward the observer located at $\theta=0$, whose radial motion, $r(t)$, is the same as the motion of the mass element of the first jet ($j=1$) in the range, 
${\rm min}\{0,\theta_v-\Delta\theta\}<\theta<\theta_v+\Delta\theta$. 
Note that this particle is not responsible for the observed GW memory even in the case of $\theta_v>\Delta\theta$.
The origin of time $T=0$ is chosen to be the arrival time at an observer of a photon emitted when the point mass starts its acceleration at $r=r_0$.
%
From this definition, the GW memory of the $j$-th thin shell is given by
\begin{widetext}
\begin{eqnarray}
    h_{\mathrm{thin}}(T,\gamma_j) 
    =\int h_\mathrm{point}(T ,\theta,\phi,\gamma_j) d\Omega
    = \frac{2 \gamma_j  \beta^2(\gamma_j) m_j G}{c^2 R \Delta \Omega} \int^{\Delta \theta + \theta_v}_{\max{[0, \Delta \theta - \theta_v]}} \frac{\sin^3 \theta \sin (2\Delta \phi)}{1 - \beta \cos \theta } \eta(T ,\theta,\gamma_j) \ d\theta,
    \label{eq:step-gw2}
\end{eqnarray}
where
\begin{equation}
    \Delta \phi =
    \begin{cases} 
        \pi : \Delta\theta > \theta_v \:\: \mathrm{and}\:\: 0 < \theta \leq \Delta\theta - \theta_v,\\ 
        \arccos\left(\cfrac{\cos\Delta\theta - \cos\theta_v \cos\theta}{\sin\theta_v \sin\theta}\right) : \mathrm{otherwise}.
    \end{cases}
\end{equation}
This formula undergoes Fourier transformation to 
\begin{equation}
    \begin{split}
    \tilde{h}_\mathrm{thin}(f, \gamma_j) &= \frac{2 \beta^2(\gamma_j) \mathcal{E}_j G}{c^4 R \Delta \Omega} \int^{\theta_v + \Delta \theta}_{\max[0, \theta_v - \Delta \theta]} \frac{\sin^3 \theta \sin (2\Delta \phi)}{1 - \beta(\gamma_j) \cos \theta } \frac{(e^{-i2\pi f T_{\mathrm{obs}}(\theta, \gamma_j)} - 1)}{(2 \pi f )^2 T_{\text{obs}}(\theta, \gamma_j)}
    e^{-i2\pi f T_s(\theta)} \ d\theta,
    \end{split}
    \label{eq:gw-thin}
\end{equation}
\end{widetext}
where the energy of the $j$-th shell is defined as
\begin{equation}
    \mathcal{E}_j = \gamma_j m_jc^2.
    \label{eq:shell-energy}
\end{equation}
The total energy of the system is written as $\mathcal{E} = \sum^N_j \mathcal{E}_j$. 
Note that the definition of Fourier transform we use is 
\begin{equation}
    \tilde{h}(f, \bm{x}) = \int^{\infty}_{-\infty} h(T, \bm{x}) e^{-i2\pi f T}dT,
\end{equation}
where $\bm{x}$ denotes the parameters that do not affect the operation of the Fourier transform.

The Fourier component of the GW memory from multiple thin shells is calculated by the superposition of the GW memory from $N$ shells, incorporating time shifts of the Fourier component for each shell as follows:
\begin{equation}
    \tilde{h}_\mathrm{thin}(f) = \sum^{N}_{j=1}
    \Big( 
    \tilde{h}_\mathrm{thin}(f, \gamma_j)
    e^{-i2\pi f \tau_j}
    \Big),
    \label{eq:gw-thin-N}
\end{equation}
where $\tau_j$ denotes the time shift from the initial accelerating shell.
Although $\tilde{h}_\mathrm{thin}(f)$ is the Fourier transform of 
$h_{\mathrm{thin}}(T,\gamma_j)$ as a function of observer time $T$,
one can find that the $j$-th shell is launched at the central engine at $t=\tau_j$. 

The characteristic strain of GW provides a useful metric to compare the sensitivities of the detectors with the detectability of GW signals~\cite{ref:moore2014gravitational}. The characteristic strain is given by $2f |\tilde{h}(f)|$, and we compute this expression in this paper.

\subsection{Single / multiple thick shells}
\begin{figure}[b]
    \centering
        \centering
        \includegraphics[width=0.6\columnwidth]{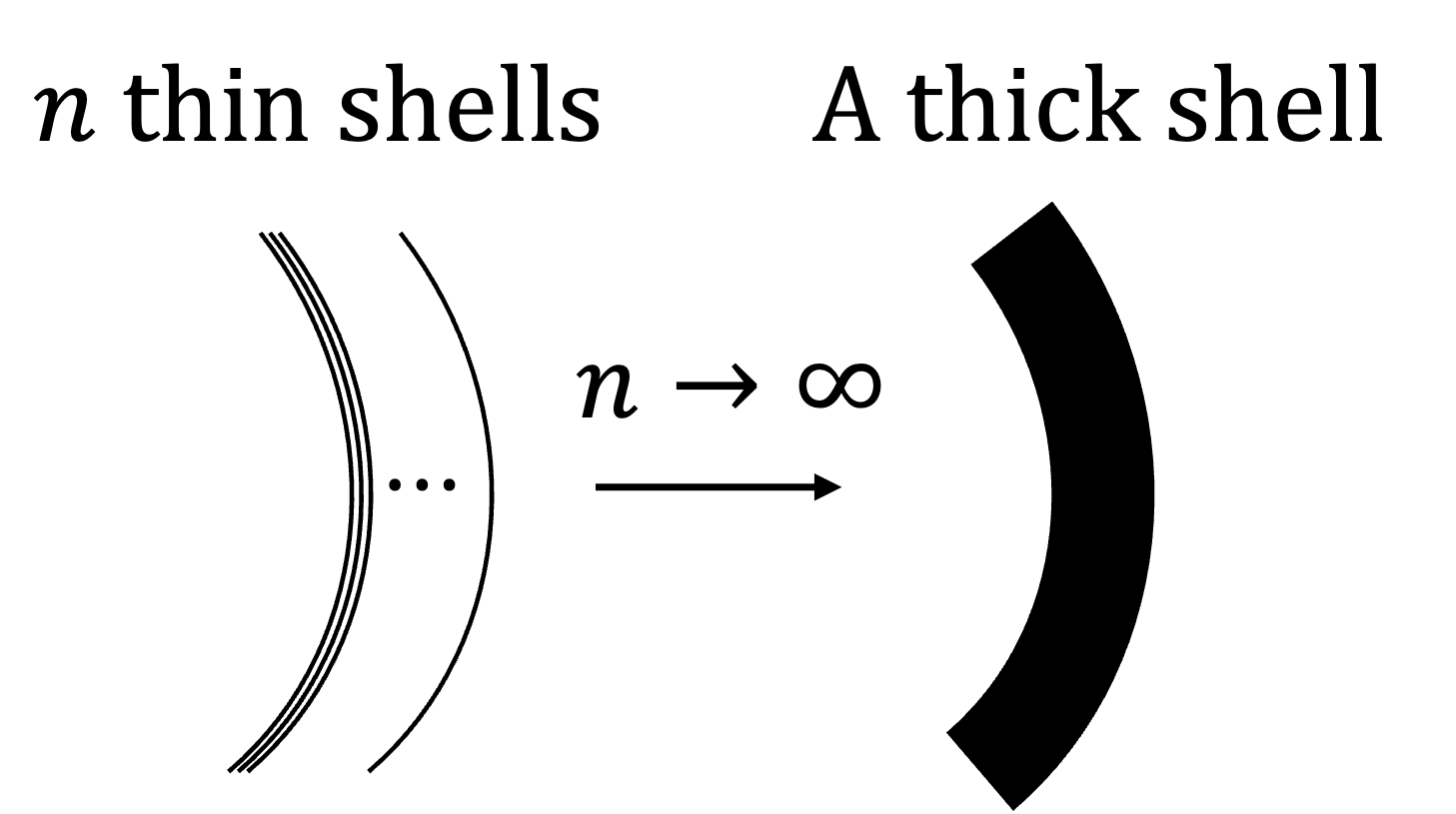}
    \caption{An illustration of a thick shell. It is assumed that the mass is uniformly distributed inside the thick shell.}
    \label{fig:nthin-thick-shell}
\end{figure}

So far, we have regarded the GW memory originating from multiple thin shells expressed as Eq.~(\ref{eq:gw-thin}) and Eq.~(\ref{eq:gw-thin-N}). Based on these equations, we extend our consideration to the case of thick shells.

Suppose that the $j$-th thick shell is composed of $n$ thin shells, 
where the thin shells are assumed to be equidistantly distributed within the thick shell. 
Unless otherwise stated, we consider the case where all thin shells have the same terminal Lorentz factor $\gamma_j$
(see Fig.~\ref{fig:nthin-thick-shell}).
We will see in Sec.~\ref{subsec:spreading-shell} that this is a good approximation.
Furthermore, let $\delta t_j$ denote the interval 
between the departure times of the front and rear ends of 
the $j$-th thick shell.
After acceleration of the whole shell, its thickness is given by
$W_j=\beta(\gamma_j)c\delta t_j$.
The thickness of the shell can be expressed as a variation 
in the time-shift term of the Fourier transform.
Let us consider a situation where thin shells are arranged with narrow spacing, with $n$ layers, which is written as:
\begin{equation}
    \tilde{h}_\mathrm{n-thin}(f, \gamma_j) = 
    \frac{1}{n}\sum^{n}_{k=0} \tilde{h}_\mathrm{thin}(f, \gamma_j) e^{-i2\pi f (\tau_j + \delta t_j \frac{k}{n})},
    \label{eq:gw-thick-small-n}
\end{equation}
where the mass is rewritten as $m = n m^{\prime}$.
We take the limit $n \to \infty$ by keeping the energy of the $j$-th thick shell constant, and
the summation is replaced by a Riemann integral (see Appendix.~\eqref{eq:piecewise-quadrature}).
The formula for a thick shell is expressed by:
\begin{equation}
    \tilde{h}_\mathrm{thick}(f, \gamma_j) = 
    \tilde{h}_\mathrm{thin}(f, \gamma_j) e^{-i2\pi f \tau_j}
    \frac{1 - e^{-i2\pi f \delta t_j}}{i2\pi f \delta t_j},
    \label{eq:gw-thick}
\end{equation}
where $\delta t_j \simeq W_j/c$ for $\gamma_j \geq 100$. 
Finally, the Fourier component of the $N$ thick shell is calculated as, 
\begin{equation}
    \tilde{h}_\mathrm{thick}(f) = 
    \sum^N_j
    \tilde{h}_\mathrm{thick}(f, \gamma_j).
    \label{eq:gw-thick-N}
\end{equation}

\red{In this paper, 
we adopt a shell thickness $W_j$ in the range, from $10^7$ to $10^9$~cm 
\cite{ref:zhang2018,ref:Piran1999PhR}.}
\red{The typical value of the
observationally estimated collimation-corrected gamma-ray energy, that is the gamma-ray emission energy confined in the jet, is about $10^{50-51}$~erg~\cite{ref:zhao2020statistical}.
Assuming a gamma-ray emission efficiency, that is, the conversion efficiency from jet kinetic energy to gamma-ray energy, of about 
5--10\%, we estimate the jet kinetic energy to be on the order, $\mathcal{E}\sim10^{52}$~erg, that is the value adopted
in the following of this paper.}

\section{Waveform properties}\label{sec:experiment}
In this section, we compute GW memories under various conditions. We investigate the differences in the representation of the GW memory waveform when the thickness $W_j$, the terminal Lorentz factor $\gamma_j$, the number of shells $N$, and the time interval between shells $\delta \tau_j=\tau_{j+1}-\tau_j$ ($j=1,\cdots, N-1$) vary. The total time in a central engine emitting GW memories is denoted as:
\begin{equation}
    T_{\mathrm{CE}} = \tau_N - \tau_1.
\end{equation}

We use the following configuration: the opening half angle and the viewing angle are set to $\Delta \theta = \theta_v = 0.1$~rad, and the starting position for acceleration is $r_0 = 10^{7}~\mathrm{cm}$, except for Sec.~\ref{subsec:single-thin-shell}. 
The distance to the source is also set to $R = 10~\mathrm{Mpc}$ in this section.

\begin{figure}[tb]
    \centering
    \begin{minipage}[t]{0.49\textwidth}
    \includegraphics[width=\columnwidth]{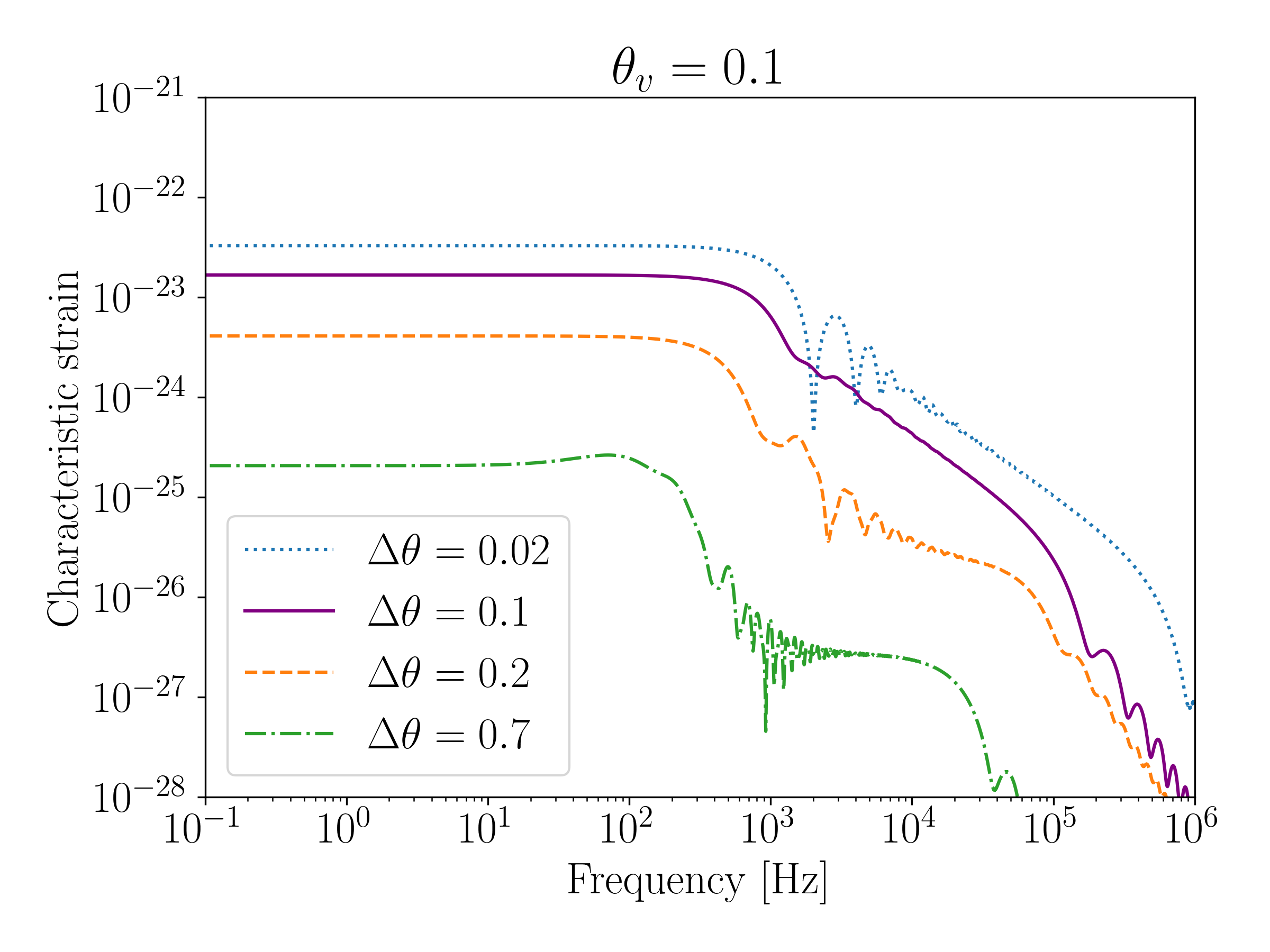}
    \subcaption{The variation in $\Delta \theta$ while keeping $\theta_v = 0.1$~rad fixed.}
    \label{fig:single-thin-shell-a}
    \end{minipage}
    \begin{minipage}[t]{0.49\textwidth}
    \includegraphics[width=\columnwidth]{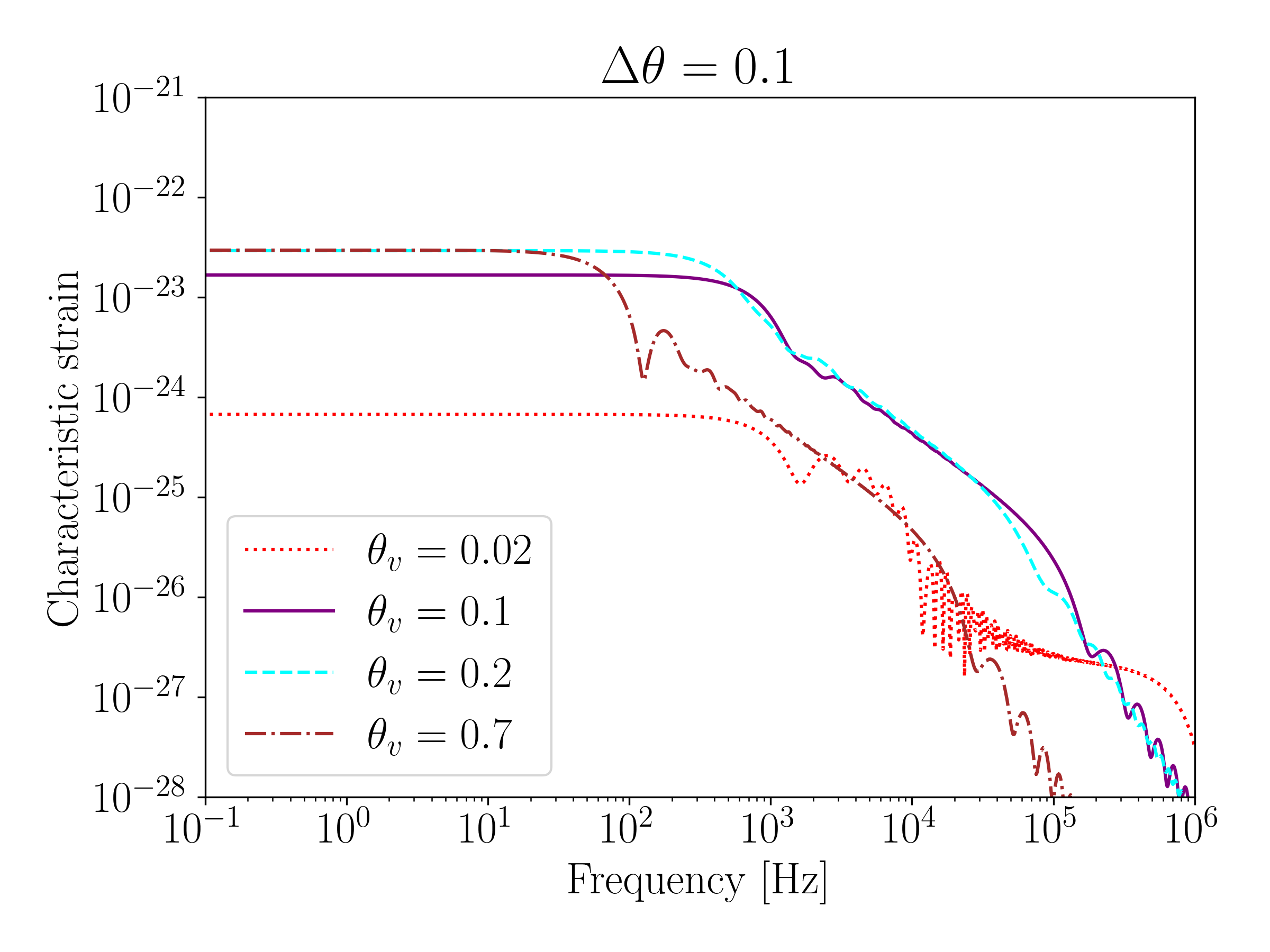}
    \subcaption{The variation in $\theta_v$ while keeping $\Delta \theta = 0.1$~rad fixed.}
    \label{fig:single-thin-shell-b}
    \end{minipage}
    \vspace{1\baselineskip}
    \caption{The Fourier waveform from a single thin shell. Each characteristic strain is given by $2 f |\tilde{h}_\mathrm{thin}(f)|$, and the total energy is set as 
    $\mathcal{E} =10^{52}~\mathrm{erg}$, $R = 10~\mathrm{Mpc}$, and $\gamma_1 = 100$ for each case.}
    \label{fig:single-thin-shell}
\end{figure}

\subsection{Single thin shell} \label{subsec:single-thin-shell}

\subsubsection{Dependency of $\Delta \theta$ and $\theta_v$}
We consider GW memory from a single thin shell using Eq.~\eqref{eq:gw-thin-N} under conditions $N = 1$ and $\gamma_1 = 100$. Figure~\ref{fig:single-thin-shell} shows the Fourier waveform with a different opening half angle $\Delta \theta$ and a viewing angle $\theta_v$. Their total energies are 
the same as $\mathcal{E} =10^{52}~\mathrm{erg}$ for comparison.
The waveform with $\theta_v = 0.1$~rad as shown in Fig.~\ref{fig:single-thin-shell-a} indicates that their amplitude depends on the opening half angle $\Delta \theta$. In the case where $\Delta \theta \leq \theta_v$, represented by the dotted blue and solid purple lines, the situation is considered where the jet's central axis is far from the observer, and the amplitude remains almost unchanged. On the other hand, for $\Delta \theta > \theta_v$, represented by the dashed orange and dot-dashed green lines, the jet's central axis gradually approaches the observer, and their amplitude decreases. 

Figure~\ref{fig:single-thin-shell-b} shows the opposite situation, where $\theta_v$ is varied while keeping $\Delta \theta$ constant. The dashed red line, $\theta_v = 0.02$, represents the case where the observer is closest to the jet's central axis, and its amplitude is the smallest. This phenomenon, in which the amplitude decreases across the entire frequency range, is known as the antibeaming effect \cite{ref:piran2002} and was demonstrated by \citeauthor{ref:sago2004}~\cite{ref:sago2004} in the context of jetted GW memory emission.
Although our waveform formula given in Eq.~\eqref{eq:gw-thin-N} includes an additional dependence on the $T_\mathrm{obs}(\theta, \gamma_j)$ term, we still confirm the presence of the antibeaming effect. 
Furthermore, our waveform for a single thin shell is consistent with previous studies\cite{ref:sago2004, ref:Birnholtz2013, ref:piran2021}.
As $\theta_v$ increases, the initial attenuation appears in the low-frequency range, such as around 100 Hz when $\theta_v = 0.7$. This is due to the contribution of $\cos\theta$ in Eq.~\eqref{eq:tobs} which becomes non-negligible. Hereafter, we set $\Delta \theta = \theta_v = 0.1$~rad in this paper to simplify the discussion.

\subsubsection{Dependency of the initial radius $r_0$} \label{subsubsec:r0}
\begin{figure}[tb]
    \centering
        \centering
        \includegraphics[width=\columnwidth]{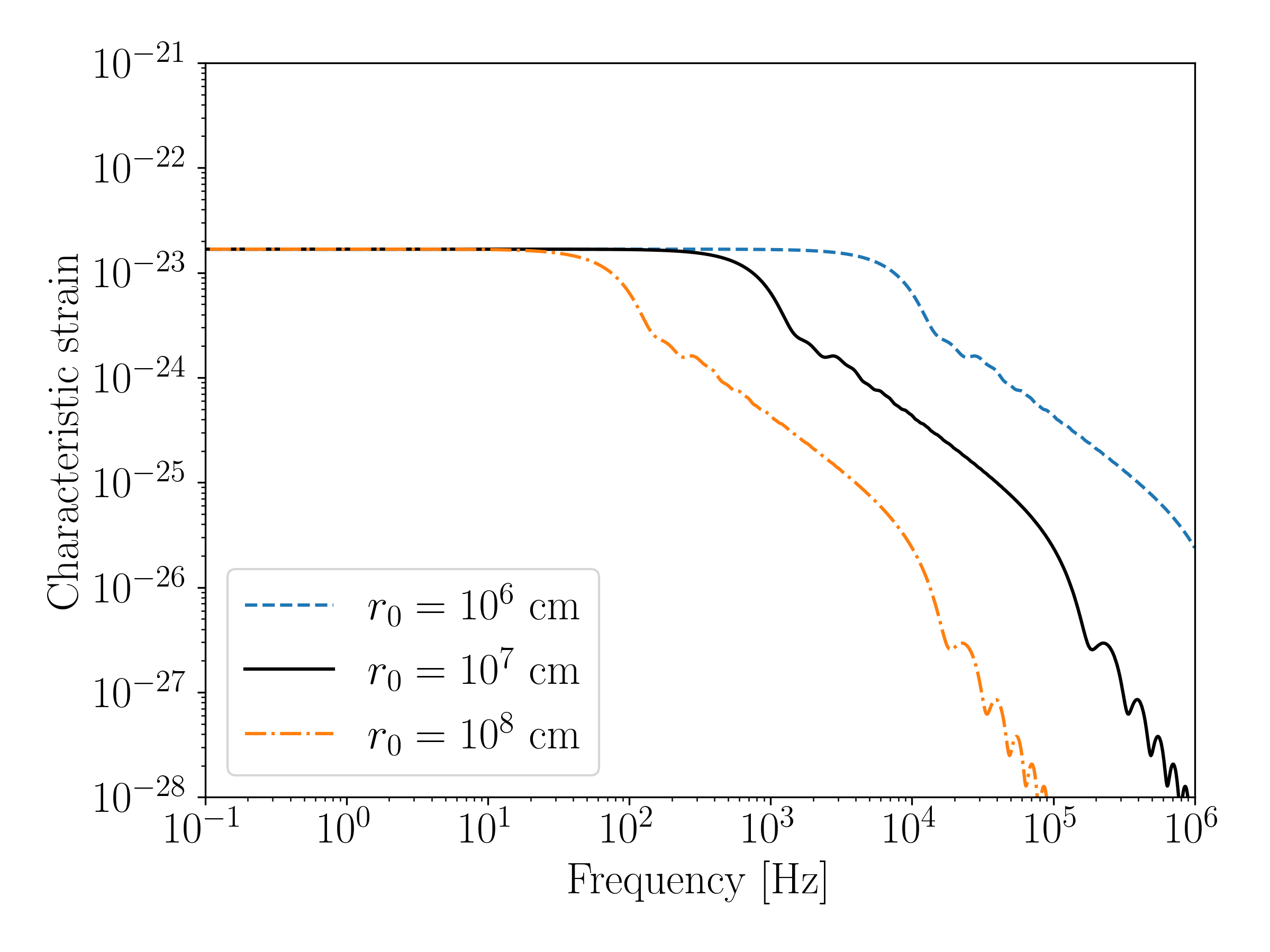}
    \caption{The waveforms for different values of the initial radius $r_0$. The solid black line, which is the same as the 
    solid purple lines 
    ($\Delta\theta = \theta_v = 0.1~\mathrm{rad}$)
    in Fig.~\ref{fig:single-thin-shell}, is plotted as a reference. 
    The total energy, the distance to the source, and the final bulk Lorentz factor are chosen as $\mathcal{E} =10^{52}~\mathrm{erg}$, $R = 10~\mathrm{Mpc}$, and $\gamma_1$ = 100, respectively.
    }
    \label{fig:acceleration-position}
\end{figure}

An order of magnitude for the observed acceleration time $T_\mathrm{obs}$, given by Eq.~\eqref{eq:order-tobs},  is mainly determined by the initial radius, $r_0$. 
We compute waveforms for several values of $r_0$ as shown in Fig.~\ref{fig:acceleration-position}. The low-frequency components at $f\ll 1$~Hz behave as the same amplitude. At low frequencies, where the second- and higher-order terms in the Taylor expansion of the exponential function are ignored, the characteristic strain in Eq.~\eqref{eq:gw-thin} is expressed as:
\begin{eqnarray}
    2 f |\tilde{h}_\mathrm{thin}(f)| &\sim& \frac{2G \mathcal{E}_1}{\pi c^4 R}\nonumber \\
     &=& 1.7\times 10^{-25}
     \left( \cfrac{\mathcal{E}}{10^{50}~\mathrm{erg}}\right)\left( \cfrac{R}{10~\mathrm{Mpc}}\right)^{-1},
     \label{eq:low-frew-approximately}
\end{eqnarray}
where we use $e^{i2\pi f T} \sim 1 + i2\pi f T$, and the integral term is approximately unity for $\gamma_1 = 100$ and $\Delta \theta = \theta_v = 0.1$~rad, as demonstrated by Fig.~1 in~\cite{ref:sago2004}. Therefore, the exponential function of $T_\mathrm{obs}$ and $T_\mathrm{s}$ vanishes at low frequencies, and the characteristic strain becomes constant at $f \ll 1$~Hz. The characteristic strain at low frequencies would be useful to estimate the total energy of GRBs because this value can be mostly estimated from $R$ and $\mathcal{E}$.

Regarding Fig.~\ref{fig:acceleration-position}, the waveform for $r_0 = 10^7~\mathrm{cm}$ at the high frequencies, we can see that the amplitude begins to attenuate around $f \sim $~kHz, because $T_{\mathrm{obs}}\sim 10^{-3}$~sec, and the factor $e^{-i2\pi f T_{\mathrm{obs}}}$ induces attenuation at these frequencies. Therefore, the frequency at which the amplitude begins to attenuate depends on $r_0$. Later, we adopt $r_0=10^7$~cm, so that $T_{\mathrm{obs}}\sim10^{-3}$~sec, and this induces a characteristic frequency, $T_{\mathrm{obs}}^{-1}\sim$~kHz, in the observed GW memory spectrum. For $\theta\approx\pi/2$, we obtain around $T_{\mathrm{obs}}\sim r_0\gamma_j/c\sim10^{-1}$~sec for $\gamma_j\sim10^2$.

\begin{figure}[tb]
    \centering
        \centering
        \includegraphics[width=\columnwidth]{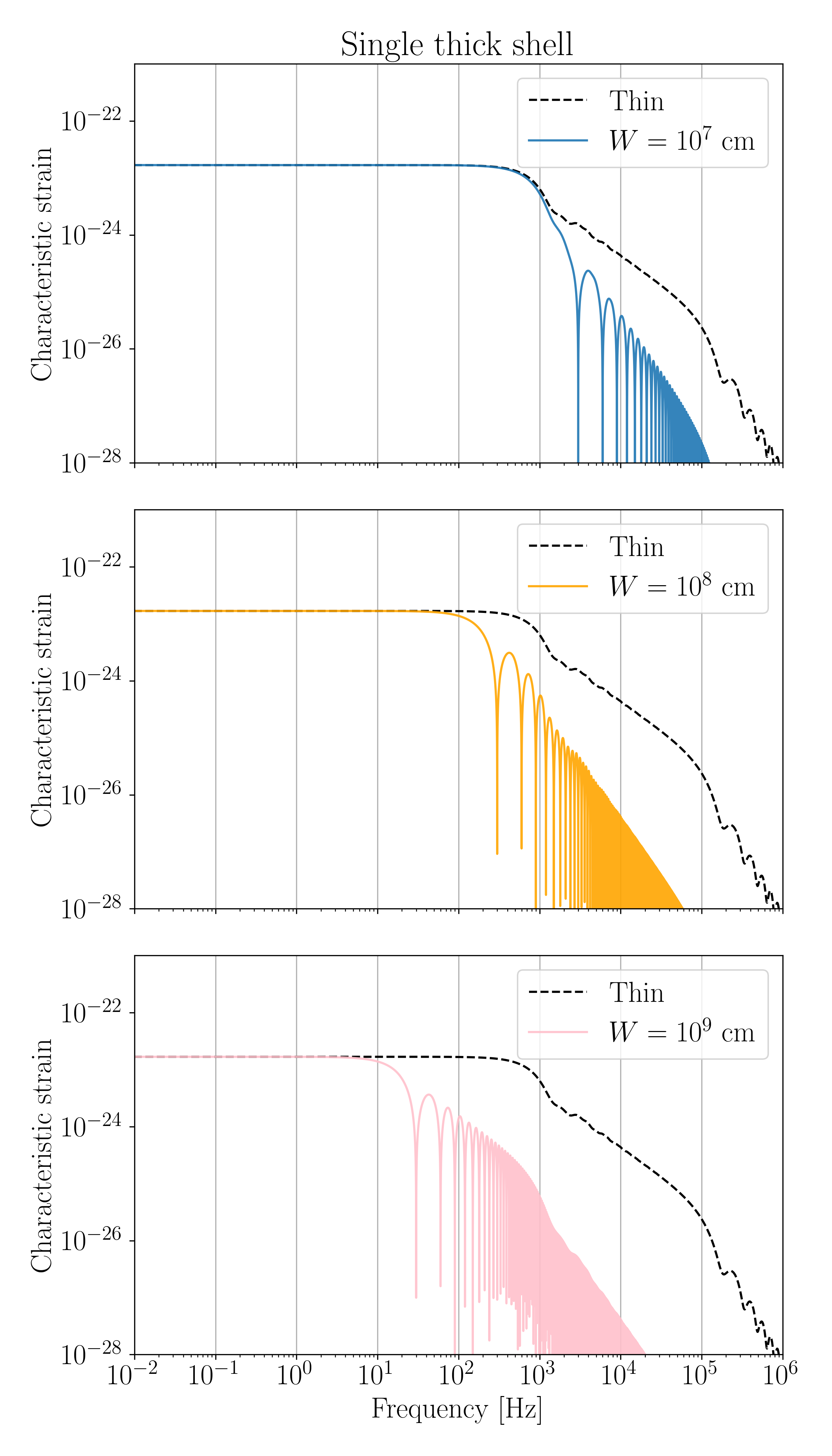}
    \caption{GW memory waveform from a single thick shell of different thicknesses. The waveform of a thin shell is also plotted with a dashed black line as a reference. 
    We set 
    $\mathcal{E} =10^{52}~\mathrm{erg}$, $R = 10~\mathrm{Mpc}$, $\gamma_1$ = 100, and $\Delta\theta = \theta_v = 0.1~\mathrm{rad}$.}
    \label{fig:single-thick-shell}
\end{figure}

\subsection{Single thick shell} \label{subsec:single-thick-shell}

\subsubsection{Single thick shell with common Lorentz factor}
We consider in this case the GW memory from a single thick shell expressed by Eq.~\eqref{eq:gw-thick-N} under situations $N=1$ and $\gamma_1 = 100$.
The waveforms of three different thicknesses, where $W_1 \simeq \delta t_1 / c$ of a single thick shell, are plotted in Fig.~\ref{fig:single-thick-shell} along with the waveform of a single thin shell for comparison. The amplitude does not change at low frequency because the waveform becomes constant due to
$e^{2\pi if \delta t_j} \sim 1 + i2\pi f \delta t_j$ in Eq.~\eqref{eq:gw-thick}, which is similar to the discussion in Sec.~\ref{subsubsec:r0}. At high frequencies, attenuation begins under the condition $f\delta t_j \geq 1$, such as when $f \sim$~kHz and $\delta t_j^{-1} \sim$~kHz. Compared to a single thin shell, high-frequency components around $f \geq $~kHz 
attenuate more than in the case of a thin shell. This attenuation is due to the effect of thickness in the term $e^{-i 2\pi f \delta t_j}$ in Eq.~\eqref{eq:gw-thick}, and we confirm that the frequency range where the attenuation begins shifts to lower frequencies as the thickness increases.

\subsubsection{Spreading of a single thick shell}
\label{subsec:spreading-shell}
\begin{figure}[htb]
    \centering
        \centering
        \includegraphics[width=\columnwidth]{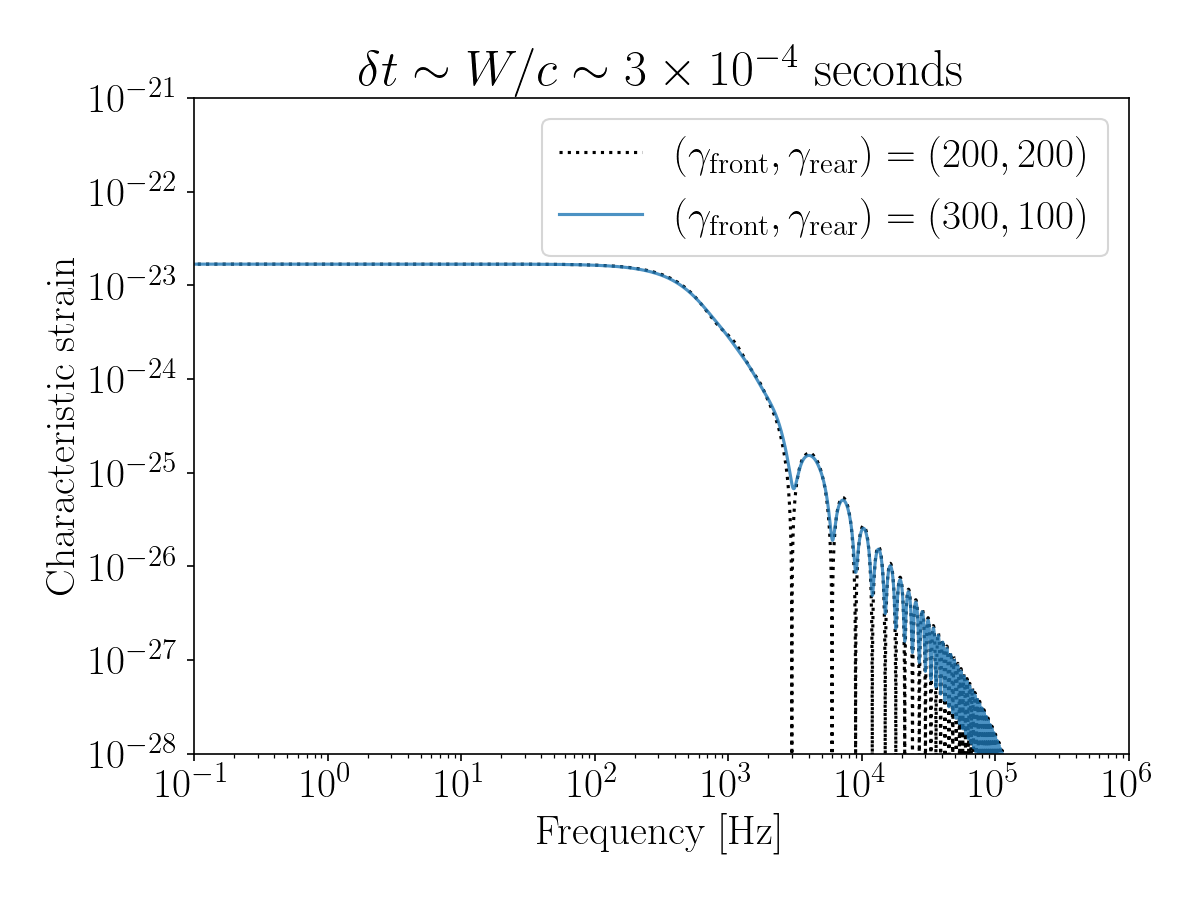}
    \caption{The waveform when the Lorentz factor at the front of a single thick shell is larger than that at the rear. The Lorentz factors of the front and rear shells follow the condition $\gamma_\mathrm{front} > \gamma_\mathrm{rear}$, which represents a thick shell spreading over time. The dashed black line shows the waveform having the same Lorentz factor for both front and rear shells. The solid blue line in Fig.~\ref{fig:single-thick-shell} is used as a reference. 
    We adopt $\mathcal{E} =10^{52}~\mathrm{erg}$, 
    $W = 10^{7}~\mathrm{cm}$, $R = 10~\mathrm{Mpc}$, $\gamma_1$ = 100, and $\Delta\theta = \theta_v = 0.1~\mathrm{rad}$.}
    \label{fig:thick-front-rear}
\end{figure}

According to the internal shock model of the prompt GRB emission, the shells spread in the radial direction because the front and rear ends of the shell have different Lorentz factors \cite{ref:zhang2018,ref:Piran1999PhR}. 
As the shells expand, their radial thickness increases. 
The shell thickness should be sufficiently large ($\sim10^{10}$~cm) when the collisions of shells occur at $r\sim10^{13-14}$~cm so that the lifetime of the internal shocks explains the rising time of the observed gamma-ray pulses. Let $\gamma_\mathrm{front}$ and $\gamma_\mathrm{rear}$ be the Lorentz factors of the front and rear shells, respectively. 
Suppose that $\gamma_\mathrm{front} > \gamma_\mathrm{rear}$ and the Lorentz factors within the single thick shell decrease monotonically from front to rear, where $\gamma_\mathrm{front}$ and $\gamma_\mathrm{rear}$ denote the maximum and minimum values in the shell, respectively.
Figure~\ref{fig:thick-front-rear} shows the waveform from a single thick shell under this condition. There is no significant difference in the waveform compared to the case that the front and rear are the same Lorentz factor, except for the sharply attenuated peak. This fact is understood as follows. The thickness of a shell just after the end of acceleration is given by
\begin{equation}
W =\beta(\gamma_\mathrm{rear})c\delta t + (1-\beta(\gamma_\mathrm{rear}))r_0\Delta\gamma,
\end{equation}
where $\Delta\gamma = \gamma_\mathrm{front} - \gamma_\mathrm{rear}$.
In the rhs of this equation, the second term is approximately $\sim r_0 / \gamma_{\mathrm{mid}} \sim 10^5~\mathrm{cm}$, which is much smaller than the first one, $\sim 10^7~\mathrm{cm}$.
Therefore, in the remainder of this paper, we neglect the effect of the shell spreading.

\subsection{Multiple thin shells: $\gamma_j$ and $\delta\tau_j$ const.} \label{subsec:multiple-thin-shells}

\begin{figure}[tb]
    \centering
        \centering
        \includegraphics[width=\columnwidth]{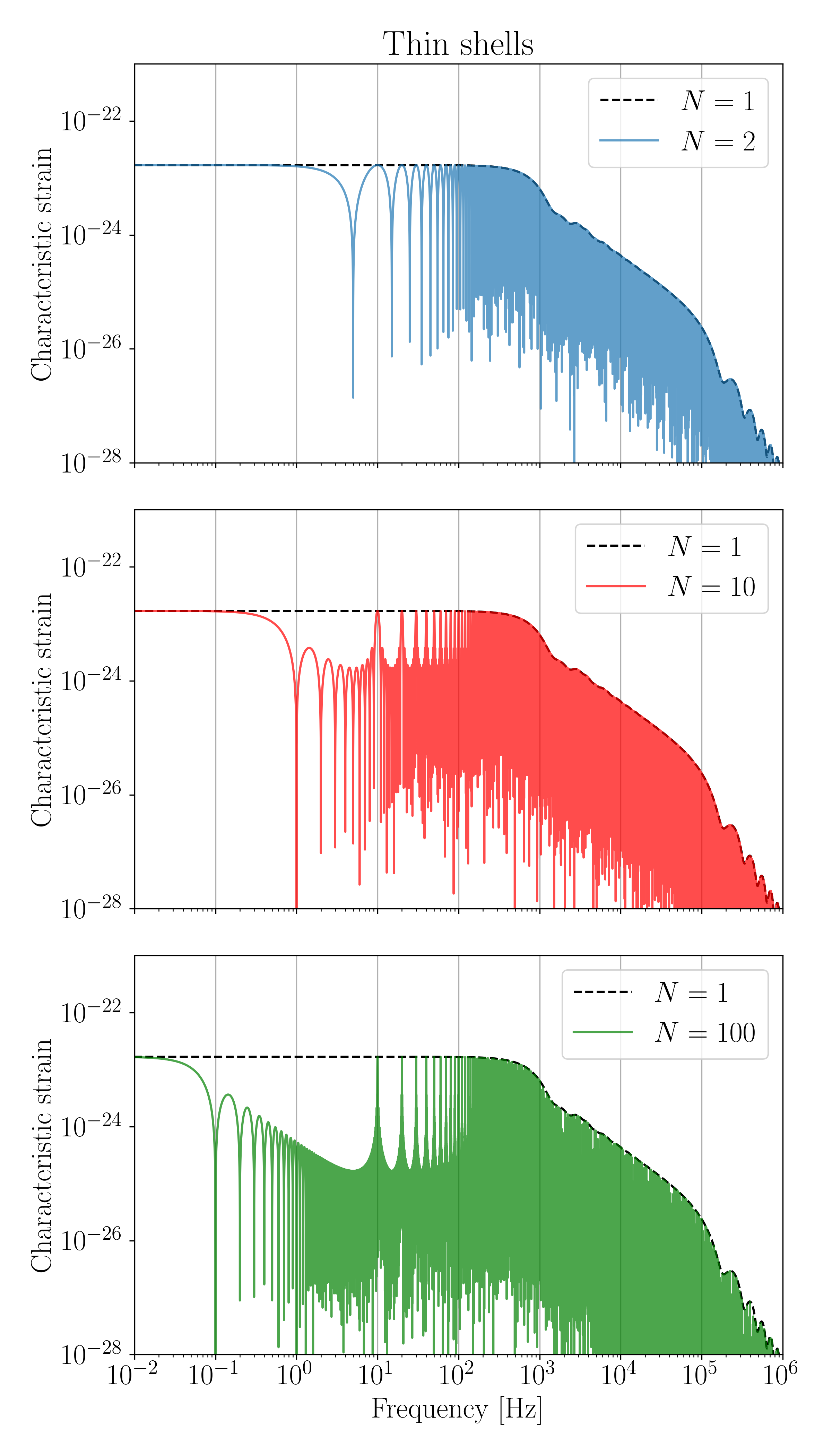}
    \caption{Characteristic strain of GW from multiple thin shells. $N$ represents the number of thin shells. $N$ shells are emitted with equal time intervals, $\delta \tau_j= 0.1$ for all $j$. The total time is $T_\mathrm{CE} = 0.1$~sec in the case of $N = 2$, $T_\mathrm{CE} \sim 1$~sec in the case of $N = 10$, and $T_\mathrm{CE} \sim 10$~sec in the case of $N = 100$. The dashed black line is plotted as a reference, which is the same as the solid purple line in Fig.~\ref{fig:single-thin-shell}. All plots have  $\mathcal{E} =10^{52}~\mathrm{erg}$, $R = 10~\mathrm{Mpc}$, $\gamma_j$ = 100 for all $j$, and $\Delta\theta = \theta_v = 0.1~\mathrm{rad}$.}
    \label{fig:multiple-thin-shells}
\end{figure}

The waveform formula from multiple thin shells is expressed in Eq.~\eqref{eq:gw-thin-N}. We plot this waveform for shell numbers $N=1,2,10,\mathrm{and}~100$ in Fig.~\ref{fig:multiple-thin-shells}. 
Each waveform is configured where the $N$ shells are 
launched with equal time intervals $\delta \tau_j = 0.1$~sec and
with the same Lorentz factor $\gamma_j = 100$ for all $j$, and $m_j$ in Eq.~\eqref{eq:shell-energy} as constant.
For comparison, the total energy of each case is set to constant as $\mathcal{E} =10^{52}~\mathrm{erg}$. The amplitude attenuation begins at low frequency under the condition $f T_\mathrm{CE}\geq 1$, such as $f \sim100$~Hz and $T_\mathrm{CE}^{-1}\sim100$~Hz in the case of $N=100$.
As a result of the time shift of the Fourier transform in Eq.~\eqref{eq:gw-thin-N}, this impact appears at lower frequencies as the number of shells increases. Therefore, this attenuation could help in understanding how many shells are formed in the central engine of the jet.

In this configuration, at frequencies 
$f = l/\delta\tau_j$, where $l$ is an integer, we can represent 
$\tau_j = 0.1 \times j$~sec under uniform time intervals, and rewrite the exponential function in Eq.(\ref{eq:gw-thin-N}) as follows:
\begin{equation}
    \tilde{h}_\mathrm{thin}(f)\Big|_{f=10l~{\rm Hz}, \tau_j = 0.1 j~{\rm sec}} = \sum^{N}_{j}
    \tilde{h}_\mathrm{thin}(f, \gamma_j)
    e^{-i2\pi j l}.
    \label{eq:same-time-interval}
\end{equation}
Because $e^{-i2\pi j \times l} = e^{-i2\pi j}= 1$ for all $j$, the exponential term does not affect the Fourier amplitude. As a result, amplitude attenuation due to the time shift does not appear 
at $f=l/\delta\tau_j=10l$~Hz,
and the peaks are observed at these frequencies for any number of shells. In other words, we can observe such peaks if the shells are emitted at uniform time intervals, though it is unnatural.
Each peak is sharper for larger $N$.
The minimum value of the peak frequency for $l=1$ is $\tau_j^{-1}=10$~Hz.
Hence, as seen in $N=100$, the observed spectrum has a sharp peak at $f=10$~Hz in Fig.~\ref{fig:multiple-thin-shells}.

\subsection{Multiple thick shells} \label{subsec:multiple-thick-shells}
We compute the waveform from multiple thick shells with several time intervals $\delta \tau_j$ and Lorentz factor $\gamma_j$. We also set $m_j$ and thickness $W_j$ as a constant.

\subsubsection{Constant $\gamma_j$ and $\delta\tau_j$}
\begin{figure}[tb]
    \centering
        \centering
        \includegraphics[width=\columnwidth]{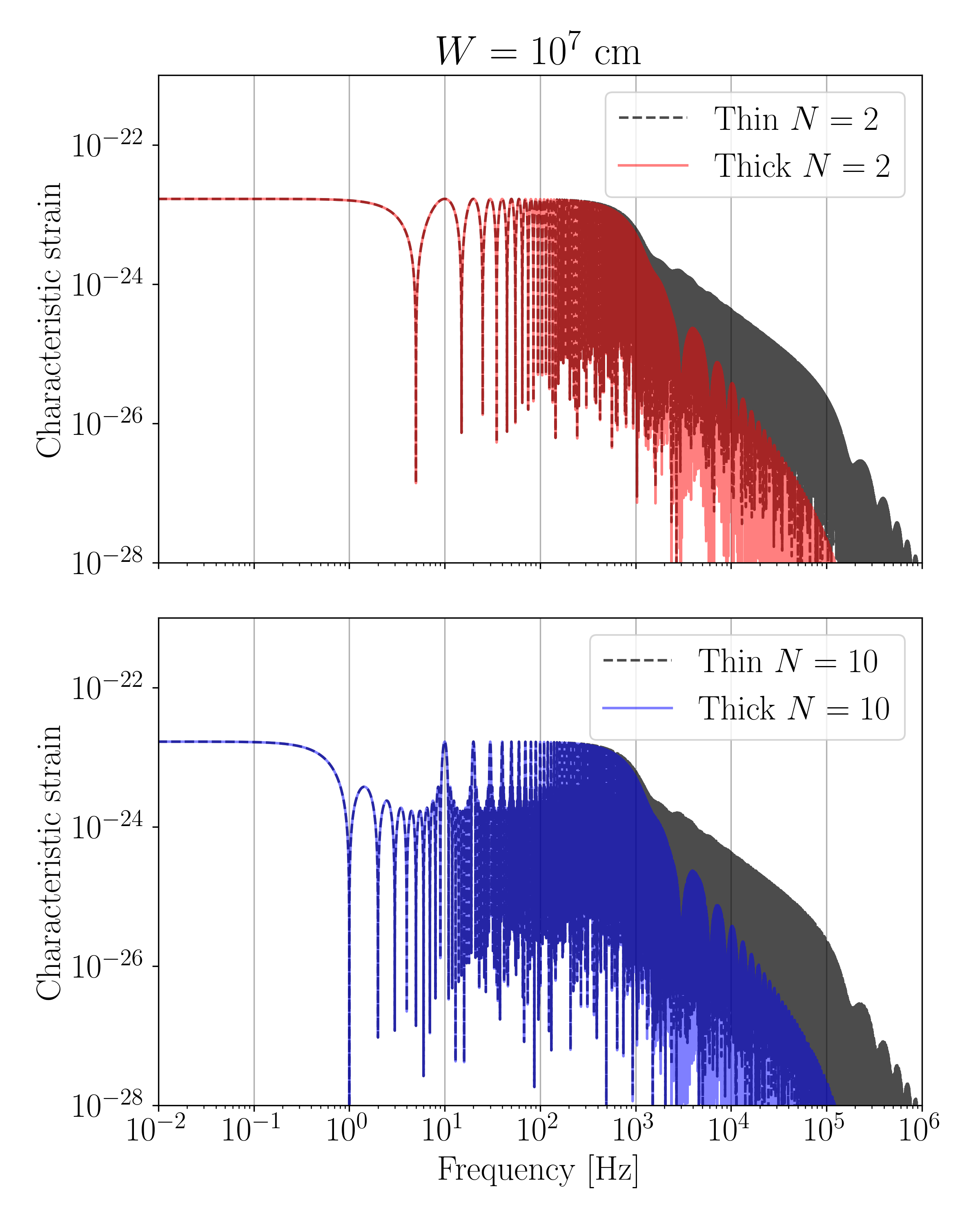}
    \caption{(a) Top: multiple thick shell model waveform, with $N=2$ being the number of shells, and the corresponding thin shell model waveform in dashed gray as a comparison. (b) Bottom: multiple thick shells, with $N=10$, and the corresponding thin shell model waveform in dashed gray as a comparison.
    All plots have $\mathcal{E} =10^{52}~\mathrm{erg}$, $W = 10^{7}\mathrm{cm}$, $R = 10~\mathrm{Mpc}$, $\gamma_j$ = 100 for all $j$, $\delta \tau_j = 0.1$ and $\Delta\theta = \theta_v = 0.1~\mathrm{rad}$.
    }
    \label{fig:multiple-thick}
\end{figure}

\begin{figure}[tb]
    \centering
        \centering
        \includegraphics[width=\columnwidth]{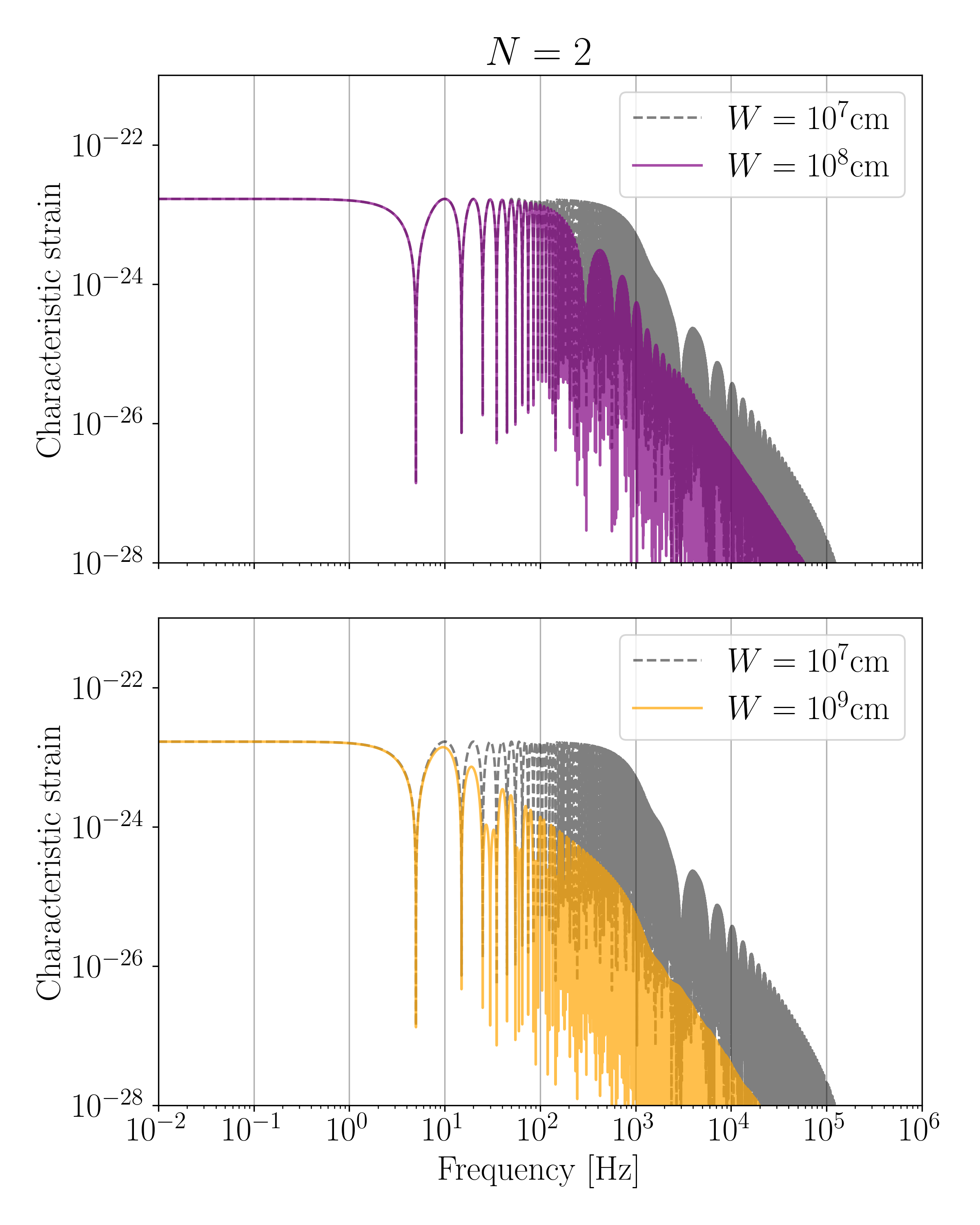}
    \caption{(a) Top: multiple thick shell model waveform, with a thickness $W = 10^{8}~\mathrm{cm}$, and the corresponding thin shell model waveform in dashed gray with $W = 10^{7}~\mathrm{cm}$ as a comparison. (b) Bottom: multiple thick shells with $W = 10^{9}~\mathrm{cm}$ and the corresponding thin shell model waveform in dashed gray as a comparison. 
    All plots have $\mathcal{E} =10^{52}~\mathrm{erg}$, 
    $N = 2$, $R = 10~\mathrm{Mpc}$, $\gamma_j$ = 100 for all $j$, $\delta \tau_j = 0.1$ and $\Delta\theta = \theta_v = 0.1~\mathrm{rad}$.}
    \label{fig:different-thickness}
\end{figure}

We compute the waveform with uniform time interval 
$\delta \tau_j = 0.1$~rad and Lorentz factor $\gamma_j = 100$ 
for all $j$.
Figure~\ref{fig:multiple-thick} shows the waveforms from $N$ thick shells along with $N = 2$ thin shells for comparison. The amplitude attenuation appears at high frequency, similarly to the discussion of Fig.~\ref{fig:single-thick-shell} in Sec.~\ref{subsec:single-thick-shell}. Such attenuation also appears at lower frequencies as the number of shells increases, which is the same feature as shown in Fig.~\ref{fig:multiple-thin-shells}. Waveforms with different thicknesses are shown in Fig.~\ref{fig:different-thickness}. In the same way as the discussion of Fig.~\ref{fig:single-thick-shell}, the frequency range where the attenuation begins shifts to lower frequencies as the thickness increases. We confirm that the effect of shell thickness and the number of shells appear independently in the waveform.

\subsubsection{Single parameter randomization: Either $\gamma_j$ or $\delta\tau_j$}

\begin{figure*}[htb]
    \centering
    \begin{tabular}{cc}
      \begin{minipage}[t]{0.49\textwidth}
        \centering
        \includegraphics[width=\textwidth]{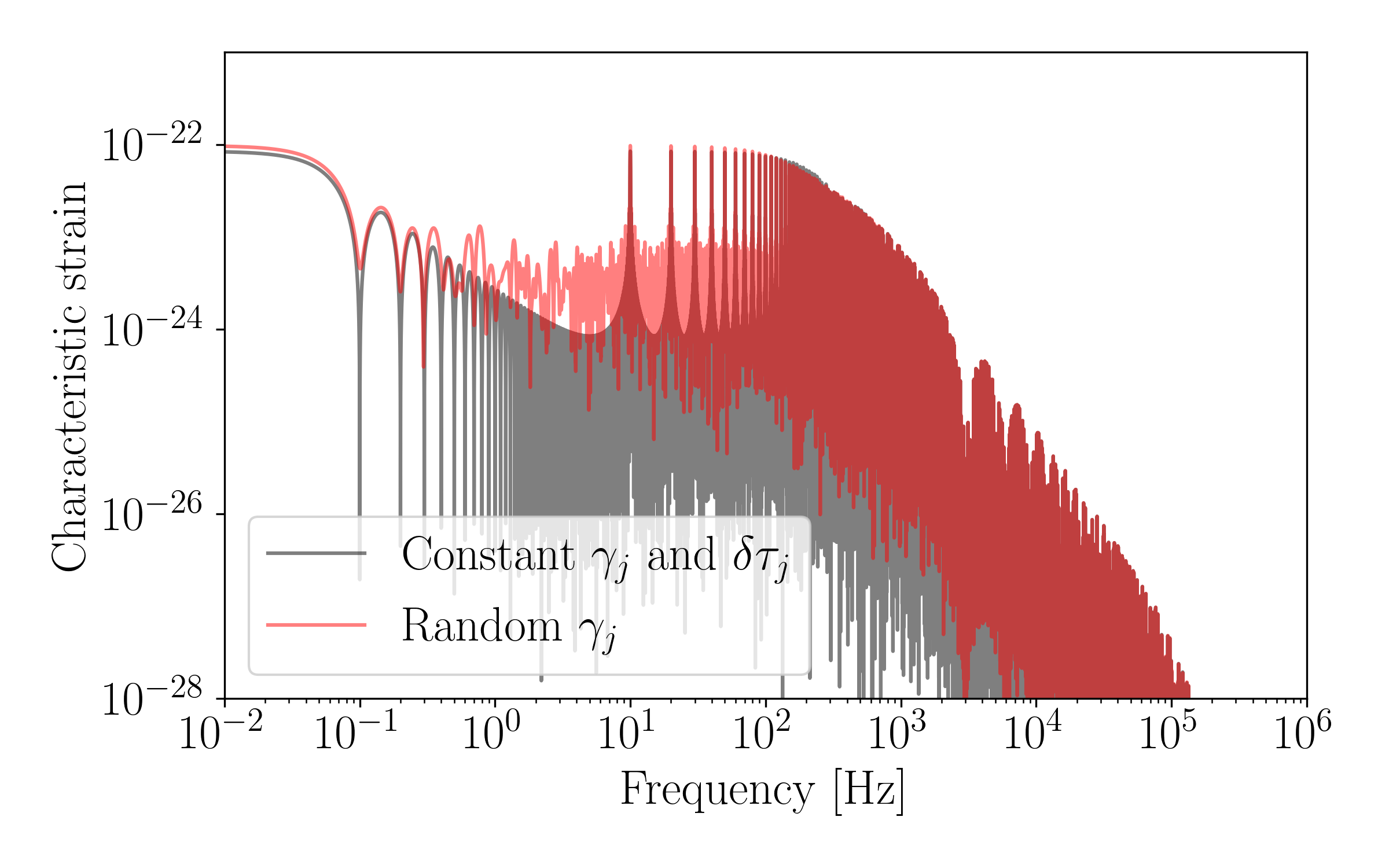}
        \subcaption{Waveform with random $\gamma_j$ and constant $\delta \tau_j$.}
        \label{fig:random-gamma-a}
      \end{minipage} &
        \begin{minipage}[t]{0.49\textwidth}
        \centering
        \includegraphics[width=0.8\textwidth]{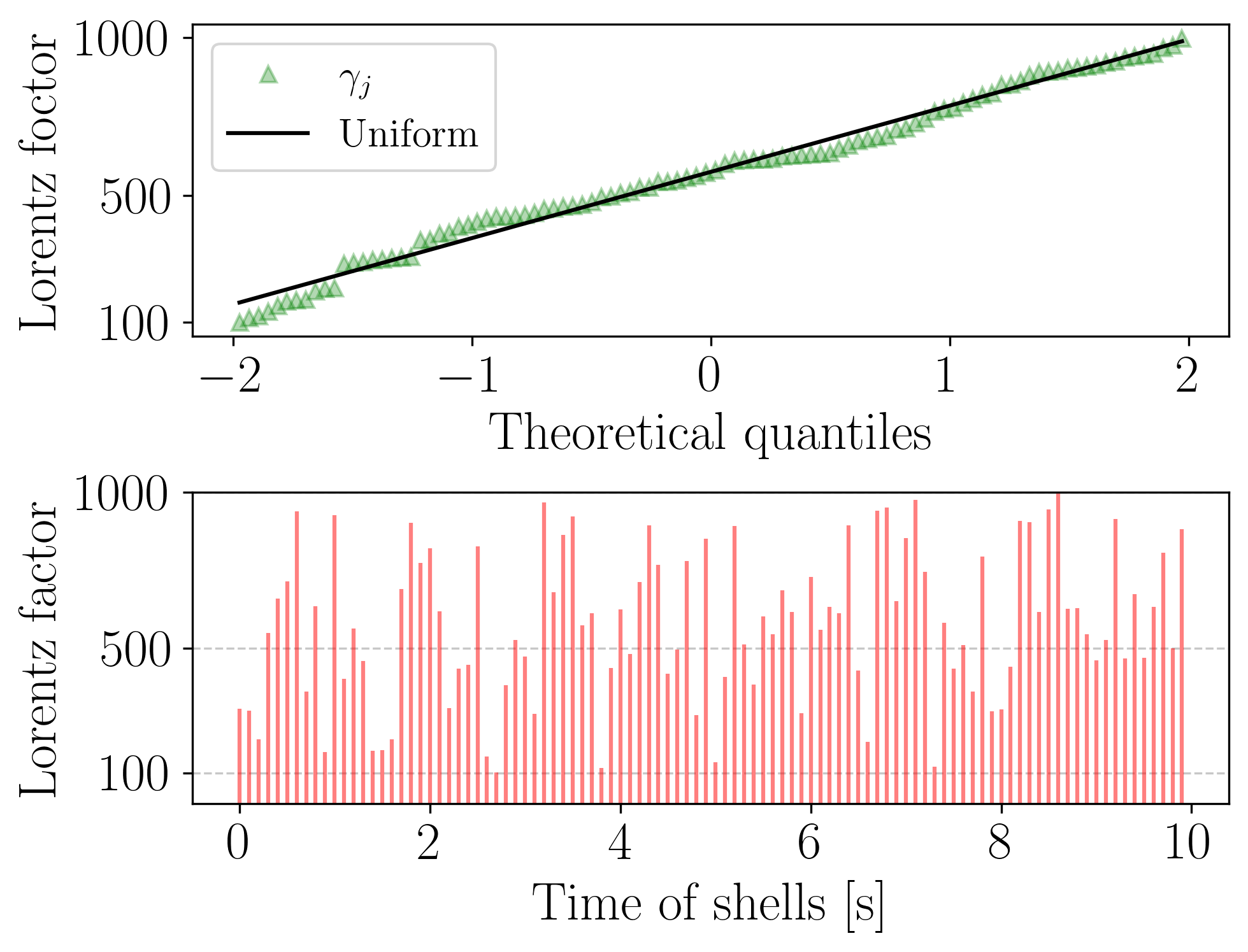}
        \subcaption{$\gamma_j$ for each shell and the corresponding Q-Q plots.}
        \label{fig:random-gamma-b}
      \end{minipage}
    \end{tabular}
    \vspace{1\baselineskip}
    \caption{(a) Left: waveform under the random $\gamma_j$. The solid gray line represents the waveform with $\gamma_j = 500$, which is the median value for the random $\gamma_j$, shown here as a reference. The corresponding energies are $\mathcal{E} = 5.75\times 10^{52}~\mathrm{erg}$ (red) and $5\times 10^{52}~\mathrm{erg}$ (gray). All plots have $W = 10^7~\mathrm{cm}$, $N = 100$, $R = 10~\mathrm{Mpc}$, $\delta \tau_j = 0.1$, and $\Delta\theta = \theta_v = 0.1~\mathrm{rad}$. \newline (b) Right: the random $\gamma_j \in [100, 1000]$ and constant $\delta\tau_j = 0.1$ are applied, and the total time is $T_\mathrm{CE} = 9.90$~sec.}
    \label{fig:random-gamma}

    \centering
    \begin{tabular}{cc}
      \begin{minipage}[t]{0.49\textwidth}
        \centering
        \includegraphics[width=\textwidth]{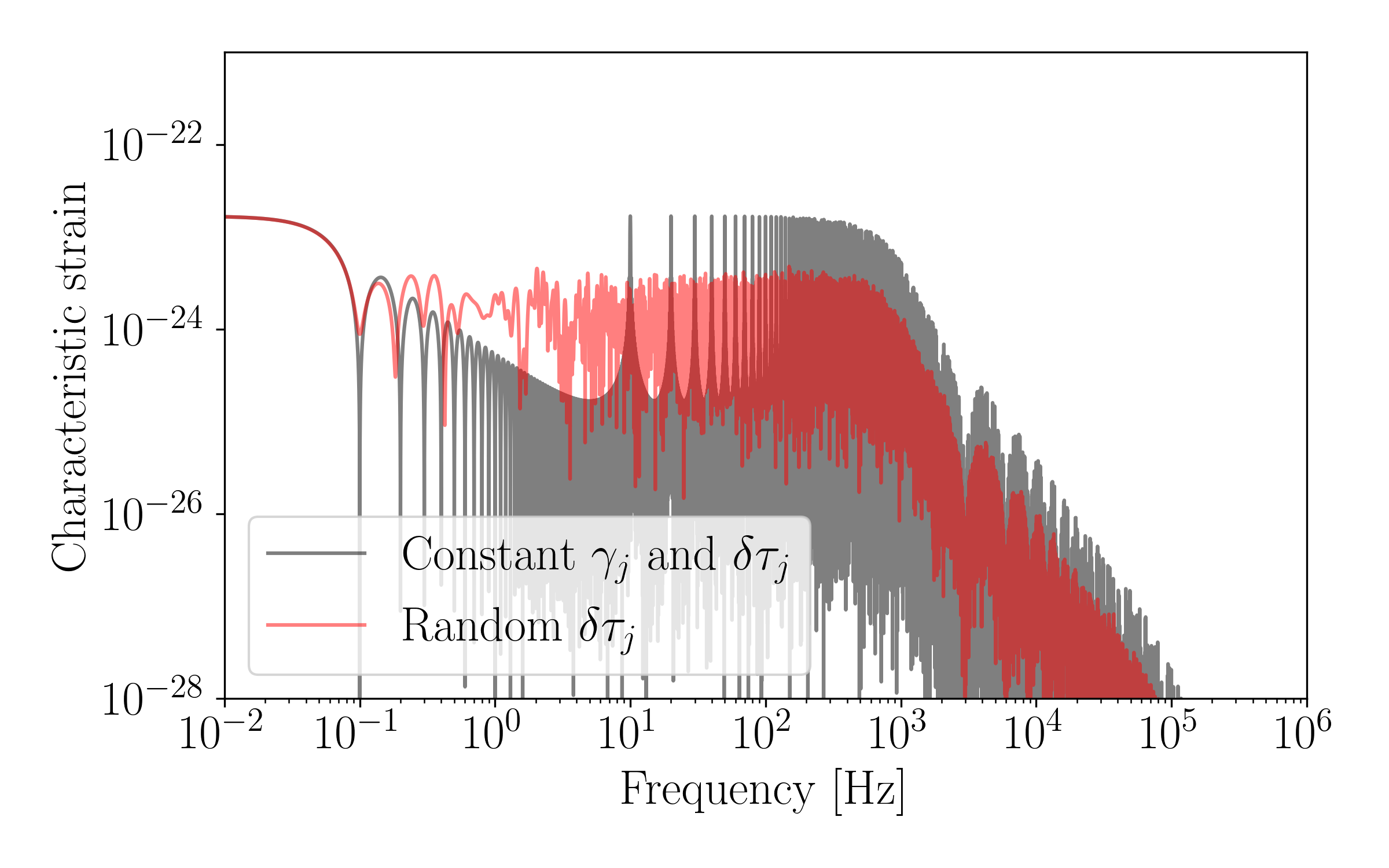}
        \subcaption{Waveform with constant $\gamma_j$ and random $\delta \tau_j$.}
        \label{fig:random-time-a}
      \end{minipage} &
        \begin{minipage}[t]{0.49\textwidth}
        \centering
        \includegraphics[width=0.8\textwidth]{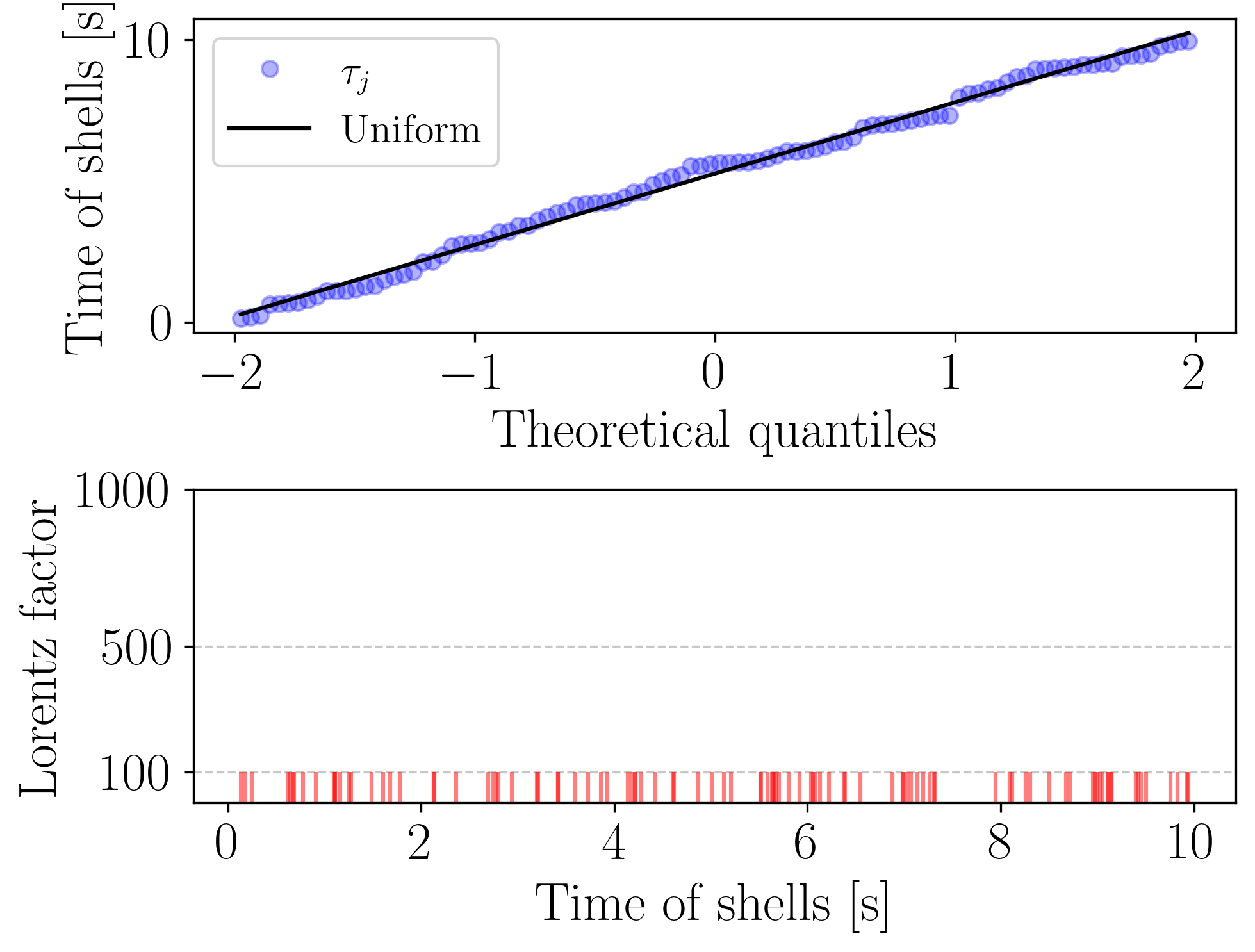}
        \subcaption{$\gamma_j$ for each shell and the corresponding Q-Q plots.}
        \label{fig:random-time-b}
      \end{minipage}
    \end{tabular}
    \vspace{1\baselineskip}
    \caption{(a) Left: waveform with random $\delta\tau_j$. The solid gray line represents the waveform with a constant $\delta\tau_j = 0.1$ as a reference. All plots have $\mathcal{E} = 10^{52}~\mathrm{erg}$, $W = 10^7~\mathrm{cm}$, $N = 100$, $R = 10~\mathrm{Mpc}$, and $\Delta\theta = \theta_v = 0.1~\mathrm{rad}$. \newline (b) Right: the constant $\gamma_j = 100$ and random $\delta\tau_j$ are applied, and the total time is $T_\mathrm{CE} = 9.81$~sec.}
    \label{fig:random-time}


    \centering
    \begin{tabular}{cc}
      \begin{minipage}[t]{0.49\textwidth}
        \centering
        \includegraphics[width=\textwidth]{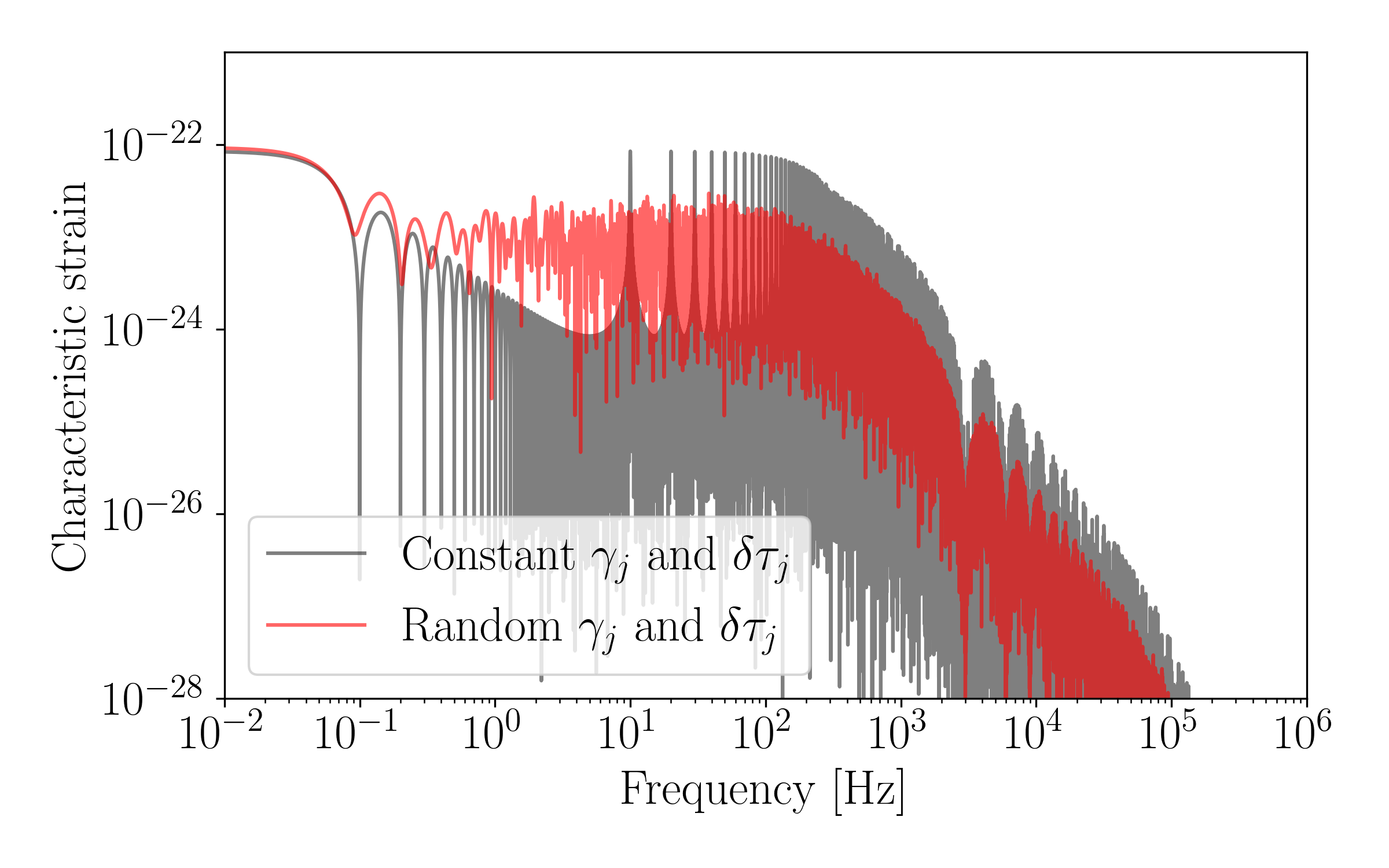}
        \subcaption{Waveform with random $\gamma_j$ and random $\delta \tau_j$.}
        \label{fig:random-gamma-time-a}
      \end{minipage} &
        \begin{minipage}[t]{0.49\textwidth}
        \centering
        \includegraphics[width=0.9\textwidth]{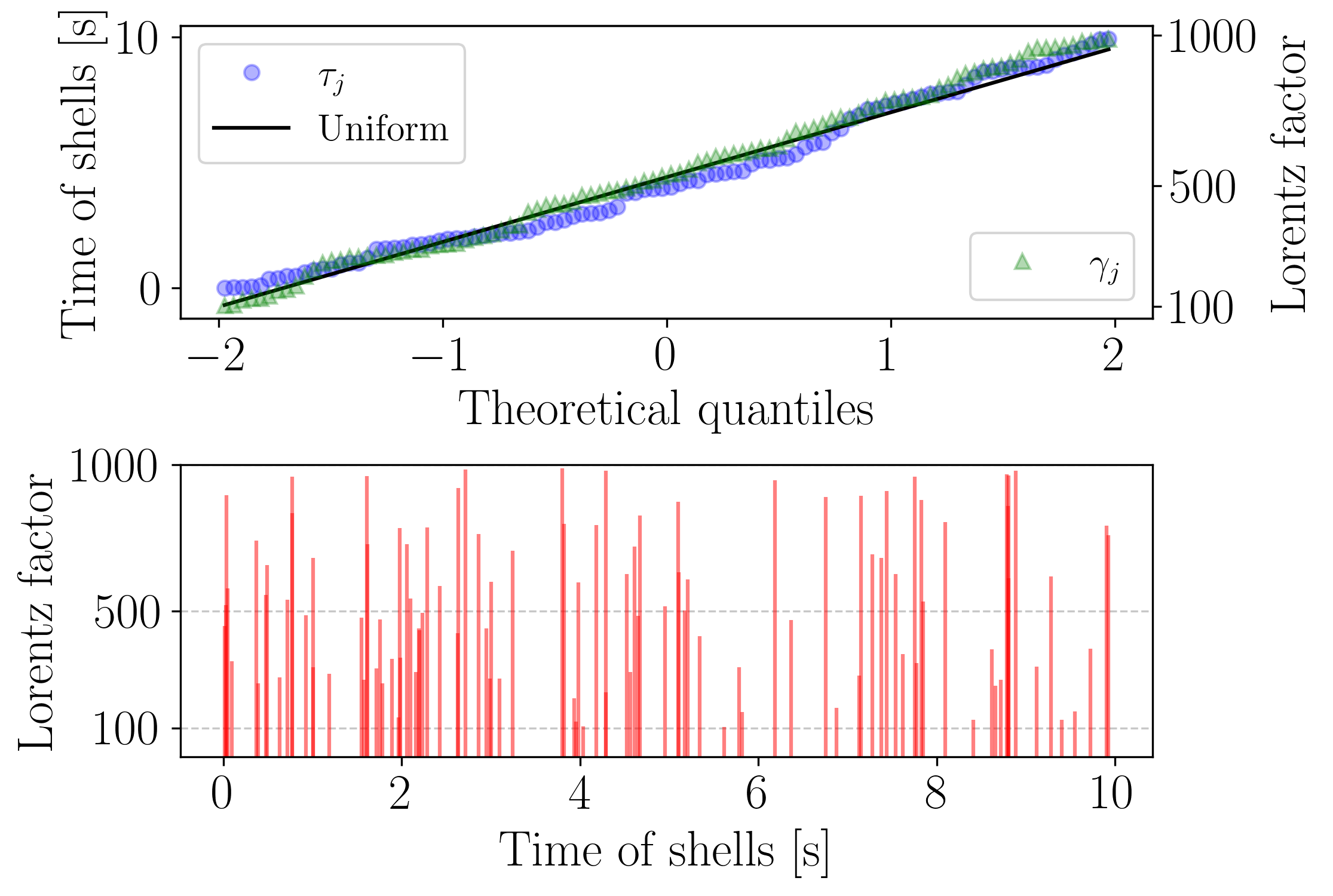}
        \subcaption{$\gamma_j$ for each shell and the corresponding Q-Q plots.}
        \label{fig:random-gamma-time-b}
      \end{minipage}
    \end{tabular}
    \vspace{1\baselineskip}
    \caption{(a) Left: waveform with random $\gamma_j$ and random $\delta \tau_j$. The solid gray line represents the waveform with constants $\delta\tau_j = 0.1$ and $\gamma_j = 500$ as a reference. The corresponding energies are $\mathcal{E} = 5.69\times 10^{52}~\mathrm{erg}$ (red) and $\mathcal{E} = 5\times 10^{52}~\mathrm{erg}$ (gray). All plots have $W = 10^7~\mathrm{cm}$, $N = 100$, $R = 10~\mathrm{Mpc}$, and $\Delta\theta = \theta_v = 0.1~\mathrm{rad}$. \newline (b) Right: the random $\gamma_j \in [100, 1000]$ and random $\delta\tau_j$ are applied for each shell, and $T_\mathrm{CE} = 9.91$~sec. A $p$ value from $\chi^2$ test is 0.911, assessing whether $\tau_j$ and $\gamma_j$ are independent.}
    \label{fig:random-gamma-time}
\end{figure*}

At the first stage and for simplicity purposes, we vary $\gamma_j$ or random $\delta\tau_j$ while keeping the other fixed to observe their individual effects on the waveform. We consider the waveforms from multiple thick shells with a random $\gamma_j \in [100, 1000]$ and a constant $\delta \tau_j = 0.1$, shown in Fig.~\ref{fig:random-gamma-a}. 
Figure~\ref{fig:random-gamma-b} shows the $\gamma_j$ assigned to each shell at each time.
The quantile–quantile (Q--Q) plot, which compares the samples to a specific distribution, indicates the sampled $\gamma_j$ follows a uniform distribution.
The solid gray line represents a constant $\gamma_j = 500$, with the total energy adjusted to be equivalent to that in the case of random $\gamma_j$. Owing to differences in $\gamma_j$, slight variations in amplitude can be seen in the frequency range $f \in [10^0, 10^2]$~Hz. In actual GRBs, constant $\gamma_j$ would prevent interactions between shells, leading to an expected waveform similar to the solid red line, while the 10 Hz peak remains visible because $\delta\tau_j$ is constant.

The waveform with constant $\gamma_j = 100$ and random $\delta\tau_j$ is shown in Fig.~\ref{fig:random-time-a}. The time intervals $\delta\tau_j$ are randomized in about 10 seconds, as shown in Fig.~\ref{fig:random-time-b}. Owing to the randomness of the time intervals, the frequency peak predicted by Eq.~(\ref{eq:same-time-interval}) is no longer present, resulting in a flattened waveform in the frequency range $f \in [10^1, 10^3]$~Hz.\\

\subsubsection{Randomization $\gamma_j$ and $\delta\tau_j$}
In this case, the Lorentz factor $\gamma_j$ and the time interval $\delta\tau_j$ of the emitted shells are both randomized to reflect a more realistic scenario. Figure~\ref{fig:random-gamma-time-a} displays the resulting waveform from multiple thick shells with randomized values of $\gamma_j$ and $\delta\tau_j$, whose values and corresponding Q--Q plots are plotted in Fig.~\ref{fig:random-gamma-time-b}.
A $p$ value from the $\chi^2$ test assesses whether $\tau_j$ and $\gamma_j$ are independent, which are sampled from a uniform distribution. The result value indicates insufficient evidence to reject the null hypothesis of independence. Therefore, we confirm that there is no correlation between them.
The waveform has a flattened shape in the frequency range around 10~Hz to 100~Hz due to the randomness $\delta\tau_j$, which is the same feature shown in Fig.~\ref{fig:random-time-a}. However, the randomness $\gamma_j$ does not affect the waveform much, and it just contributes to the total energy expressed in 
Eq.~(\ref{eq:low-frew-approximately}).\\

We demonstrate the waveform implementations of GW memory 
under various configurations: 
examining the dependency on the opening half angle $\Delta\theta$,
the viewing angle $\theta_v$,
and the initial radius $r_0$, waveforms from multiple thin and thick shells, and variations in final Lorentz factors and time intervals. Distinct waveform shapes are confirmed across different setups.

\clearpage 
\section{Detectability in quasirealistic scenarios}\label{sec:practical}
In this section, we examine the detectability of more realistic GW memory by the next-generation detectors, ET and DECIGO. The detector overviews are described in  Appendix~\ref{app:next-gw-detectors}. As part of this realistic scenario, both the terminal Lorentz factor $\gamma_j$ and the time interval of emitted shells $\delta\tau_j$ are supposed to be random.

\subsection{Waveforms with typical $T_\mathrm{CE}$}
\begin{figure}[b]
    \centering
        \centering
        \includegraphics[width=\columnwidth]{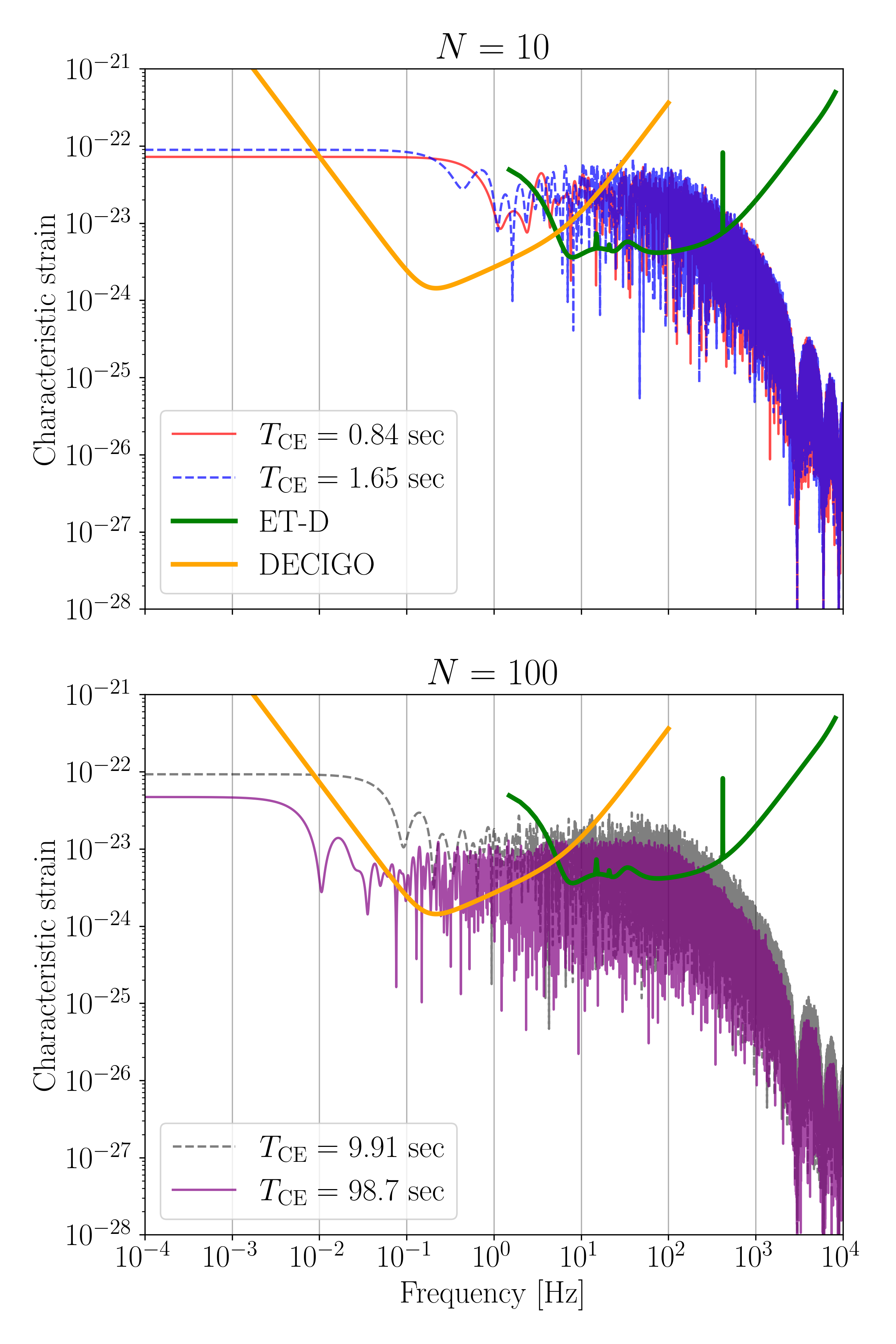}
    \caption{The waveforms for different $T_\mathrm{CE}$. All plots have $W = 10^7\mathrm{cm}$, $R = 10~\mathrm{Mpc}$, the random $\gamma_j \in [100, 1000]$ and $\delta \tau_j$ are set for each shell, and $\Delta \theta = \theta_v = 0.1$~rad.
    Top:
    the case of $T_\mathrm{CE} = 0.84$ and 1.65~sec has the total energy $\mathcal{E} = 4.27\times 10^{52}~\mathrm{erg}$ and $\mathcal{E} = 5.27\times 10^{52}~\mathrm{erg}$, respectively.
    Bottom:
    the solid purple line shows the case of $T_\mathrm{CE} = 98.7$~sec with $\mathcal{E} = 2.78\times 10^{52}~\mathrm{erg}$.
    The dashed black line is also plotted as a reference, which is the same waveform as in Fig.~\ref{fig:random-gamma-time-a}.
    }
    \label{fig:total-times}
\end{figure}

Based on actual GRB observations~\cite{ref:Guetta2004}, e.g., in the Burst And Transient Source Experiment (BATSE) catalog~\cite{ref:paciesas1999fourth}, 
the value of $T_\mathrm{CE}$ ranges
from a few seconds to a few hundred seconds.
Therefore, we compute waveforms with different $T_\mathrm{CE}$ shown in Fig.~\ref{fig:total-times} along with the DECIGO and ET-D sensitivity curves. All waveforms are computed with randomly assigned $\gamma_j$ and $\delta\tau_j$ parameters to better reflect realistic conditions.
The waveforms lie on DECIGO’s sensitivity curve in the low-frequency range, on ET-D’s curve in the high-frequency range, and around 10 Hz, it overlaps with both sensitivity curves. This result suggests the possibility of simultaneously observing the GW memory from GRBs with both ET-D and DECIGO.
The longer the $T_\mathrm{CE}$, the more the initial attenuation shifts toward the lower frequency band, which would be observable in the $\leq 1$~Hz range by DECIGO.
As discussed in Sec.~\ref{sec:experiment}, the attenuation amplitude over the frequency range depends on the physical configuration of the GRBs. Therefore, a key step toward further understanding of GRBs is to observe their GW memories using multiple detectors that cover a wide range of frequency bands.

\subsection{Nearby GRB~221009A-like event}

An energetic event GRB~221009A occurred nearby 
(redshift $z=0.151$, corresponding to the luminosity distance 
of 724~Mpc) \cite{ref:burns2023grb,ref:Frederiks2023grb}.
Its bright observed flux
in the prompt GRB emission phase reaches Earth 
every $\sim10^4$~yr.
The isotropic equivalent gamma-ray energy of this event was
about $E_{{\rm iso,}\gamma}\simeq1.2\times10^{55}$~erg.
Well-sampled afterglows, spanning from radio to very-high-energy gamma rays, suggest that the event was associated with a narrowly collimated jet characterized by an opening half angle of $\Delta\theta\sim0.01$~rad or smaller
\cite{ref:Cao2023,ref:Laskar2023,ref:Sato2023grb221009A,
ref:Zhang2023,ref:Ren2024,ref:Zheng2024,ref:derishev2024}.
Then, the collimation-corrected energy of the gamma-ray emission
is estimated as 
$(\Delta\theta)^2E_{{\rm iso,}\gamma}/4\simeq3\times10^{50}$~erg.
If we assume the efficiency of the prompt GRB
emission of 10\%, the jet kinetic energy may be
$\mathcal{E} \sim 3\times 10^{51}~\mathrm{erg}$.

If similar events to GRB~221009A occurred very nearby, then 
the GW memory is detectable.
Because the light curve of 
GRB~221009A consists of roughly five pulses in $\sim510$~seconds~\cite{ref:Frederiks2023grb},
we adopt $N=5$ and total time $T_\mathrm{CE} \sim 510$~s.
The distance to the source is assumed to be approximately one twentieth that of GRB~221009A, i.e., $R=30$~Mpc. \\

\begin{figure*}[htb]
    \centering
    \begin{tabular}{cc}
      \begin{minipage}[t]{0.49\textwidth}
        \centering
        \includegraphics[width=\textwidth]{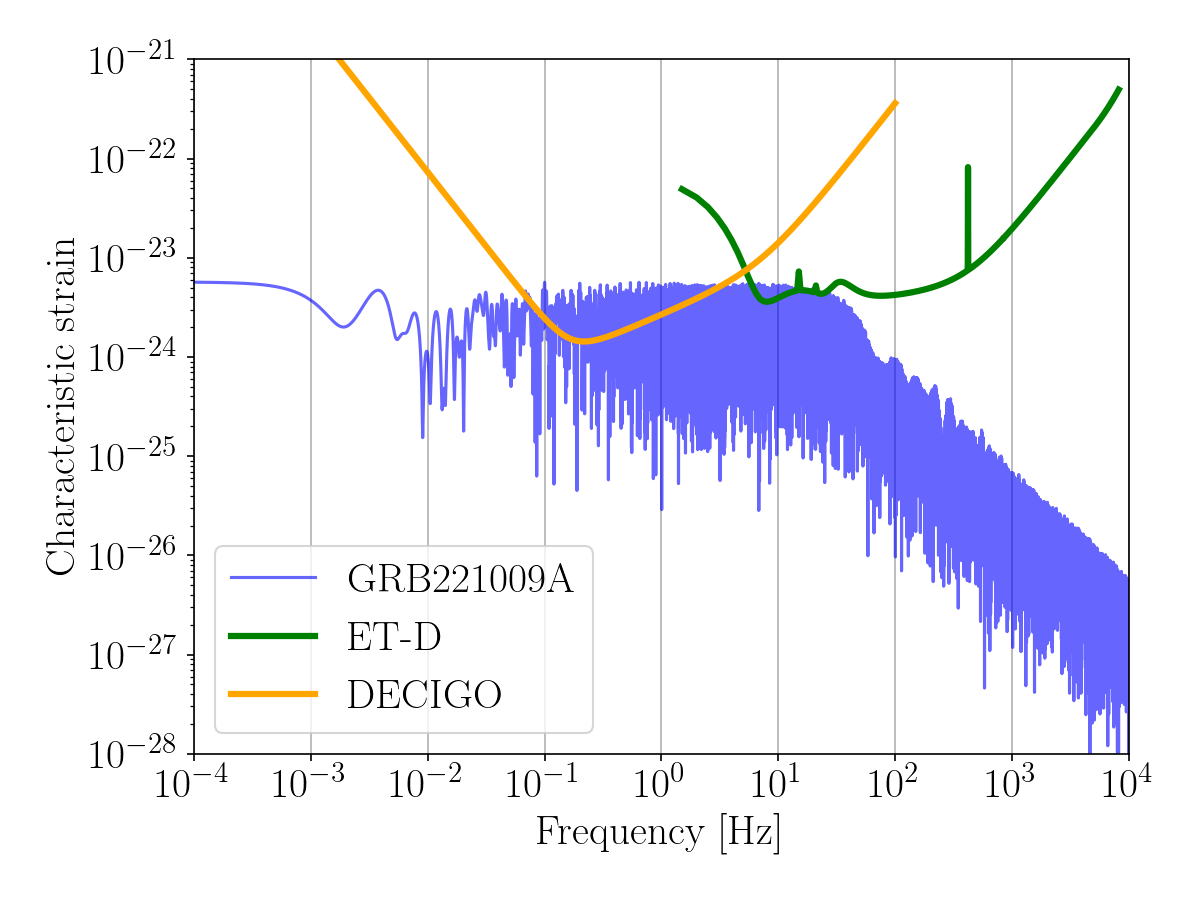}
      \end{minipage} &
        \begin{minipage}[t]{0.49\textwidth}
        \centering
        \includegraphics[width=\textwidth]{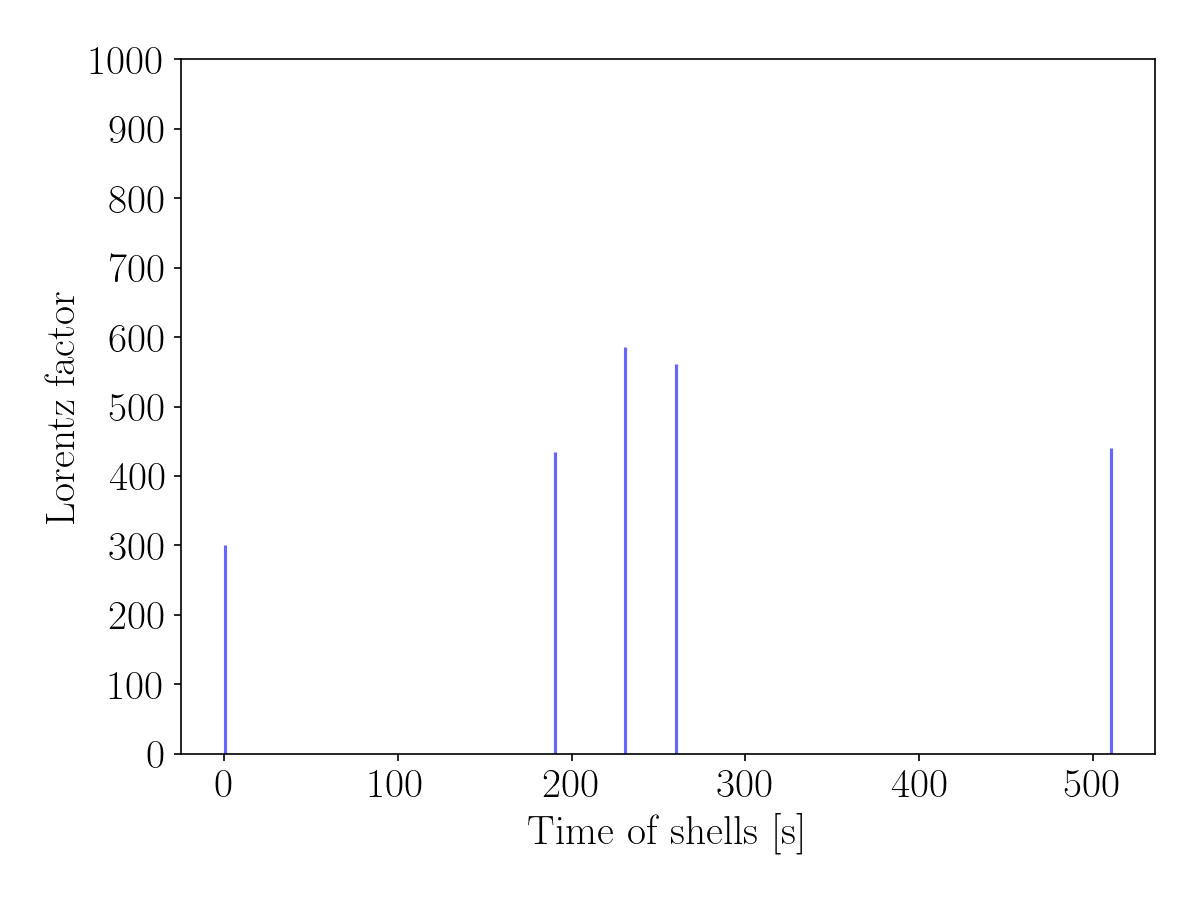}
      \end{minipage}
    \end{tabular}
    \caption{Hypothetical waveform from nearby GRB~221009A event. The distance to the source is $R=20$~Mpc. We use $\mathcal{E} = 3.48\times 10^{51}~\mathrm{erg}$, the random $\gamma_j \in [300, 600]$ and $\delta \tau_j$ are applied for each shell, $r_0 = 10^7~\mathrm{cm}$, $T_\mathrm{CE}$ = 510~sec, $N = 5$, and $W = 10^6~\mathrm{cm}$, $\Delta\theta = 0.01$ rad, and $\theta_v = 0.4$ rad.}
    \label{fig:GRB221009A}
\end{figure*}

The waveform is plotted in Fig.~\ref{fig:GRB221009A}. 
It shows the detectability to observe the GW memory around 10~Hz and 100~Hz by ET-D. The attenuation at low frequency begins from $\sim 10^{-3}$~Hz because $fT_\mathrm{CE} \geq 1$ with $T_\mathrm{CE}^{-1} \sim 10^{-3}$~Hz, which are the exponents of exponential function in Eq.~\eqref{eq:gw-thick}.
The attenuation due to the shell thickness should appear at the frequency $\delta t_j^{-1} \sim (W_j/c)^{-1}\sim10^{4}$~Hz; however, those amplitudes are already below the sensitivity curve of ET-D. 
This is because the attenuation associated with $T_\mathrm{obs}^{-1} \sim 10^2~$Hz begins earlier than that of $\delta t_j$ under the conditions $\gamma \in [300, 600]$, $r_0 = 10^7~\mathrm{cm}$, and $\theta \in [0, 0.4]$.
The characteristic strain determined by Eq.~\eqref{eq:low-frew-approximately} at low frequency is $\sim 10^{-23}$, which is observed around $10^{-4}$~Hz in Fig.~\ref{fig:GRB221009A}. Furthermore, because the magnitude of characteristic strain is almost the same as that observed in the ET-D around $10^{2}$~Hz and is determined by $R$ and $\mathcal{E}$, it may contribute to the estimation of $\mathcal{E}$ of the GRBs with only ET-D observations.\\

{\red{For the parameters given above, we find that}}
the maximum detectable distance to the source is found to be approximately $\sim20$~Mpc for ET-D and $\sim60$~Mpc for DECIGO. If the distance is about 20~Mpc, our hypothetical event is simultaneously detectable by both detectors. 
Assuming that GW memory from GRB~221009A-like jets is detectable up to a distance of 60~Mpc, the likelihood of observing an on axis jet is extremely low: about one event every $\sim2\times10^7$~yrs.
However, owing to the anti-beaming effect of GW memory emission, the jet's GW signal is detectable even for off axis viewing angles. With a beaming factor of
$f_b=2\pi/\Delta\Omega\sim2\times10^4$, the detection rate is boosted to roughly one event every $\sim10^3$~yrs.
{\red{
This estimate neglects the dependence of the signal amplitude 
on the viewing angle $\theta_v$. Furthermore, the opening angle
of the core of the jet of GRB~221009A is currently still uncertain.
Hence our estimation on the rate would be greatly changed
\cite{ref:Cao2023,ref:Laskar2023,ref:Sato2023grb221009A,
ref:Zhang2023,ref:Ren2024,ref:Zheng2024,ref:derishev2024}.
}}
%
%
Although the detection rate of such events may remain low, nearby events with GW memory could provide valuable insights into the jet launching mechanism.

\begin{figure}[htb]
    \centering
        \centering
        \includegraphics[width=\columnwidth]{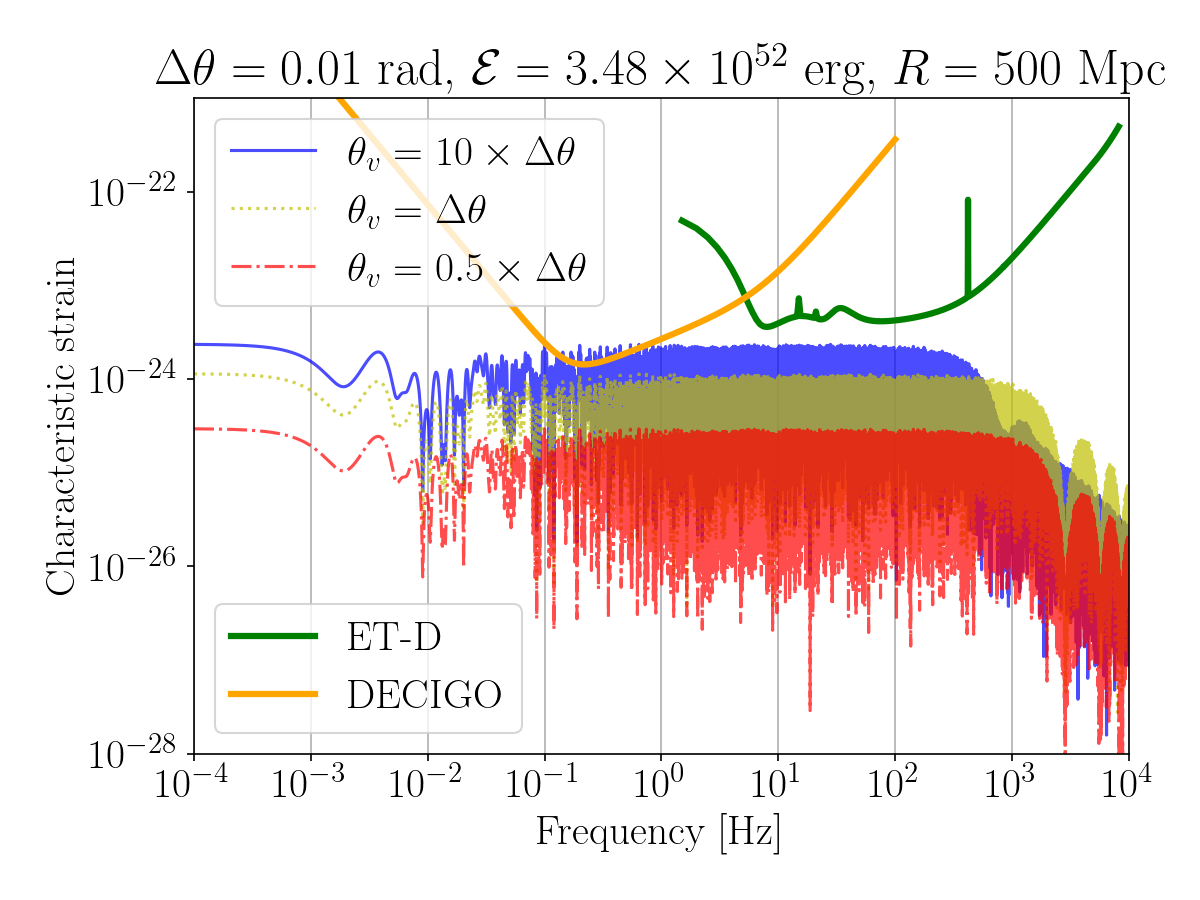}
    \caption{The waveform, using the same parameters as in Fig.~\ref{fig:GRB221009A}, except for the distance to the source $R$, the total energy $\mathcal{E}$, and the combination of angles $\theta_v$ and $\Delta \theta$.}
    \label{fig:GRB221009A-500Mpc}
\end{figure}

\red{Let us suppose a more energetic event at a larger distance, such as $\mathcal{E}\sim 10^{52}$~erg and $R = 500$~Mpc. The resulting waveforms for various combinations of angles $\theta_v$ and $\Delta\theta$ are shown in Fig.~\ref{fig:GRB221009A-500Mpc}.
Even at such a large distance, the GW memory is still detectable by DECIGO when observed off-axis, i.e., in the case of $\theta_v = 10 \times \Delta\theta$, which is roughly consistent 
with the result given in Ref.~\cite{ref:piran2022arXiv}. 
Note that in this paper, the effects of cosmic expansion are not taken into account. However, the luminosity distance of 500~Mpc corresponds to a redshift $z\sim0.1$. Hence, they do not still qualitatively alter our argument.}

\subsection{Possible GW emission from a black hole formation event in M31}
%
Recently, the disappearance of a massive star in M31 was observed, marking the formation of a black hole~\cite{ref:De2024}.
The progenitor star, with an initial mass of approximately $20M_\odot$, evolved into a red supergiant before collapsing. At the terminal nuclear burning phase, its mass was estimated at $6.7M_\odot$ , and the star’s core collapsed and turned into a black hole with a mass $6.5M_\odot$. During its core collapse, a part of the stellar material was ejected. One can expect that some of it will likely generate a relativistic jet. However, the lack of bright optical emission suggested that, even if such a jet was launched, it likely did not escape from the star. If the GW memory had been detected, we would have obtained evidence of the jet in the star.

\begin{figure*}[htb]
    \centering
    \begin{tabular}{cc}
      \begin{minipage}[t]{0.49\textwidth}
        \centering
        \includegraphics[width=\textwidth]{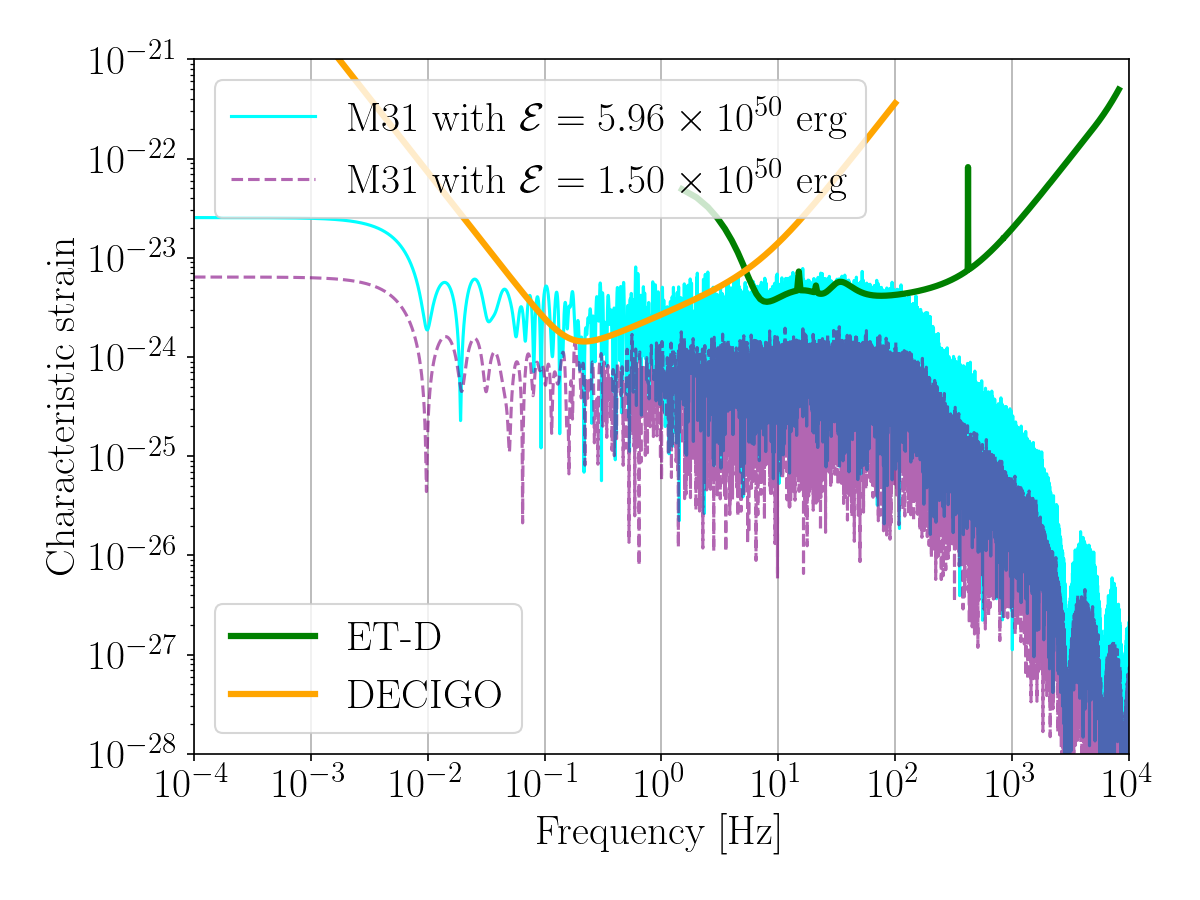}
      \end{minipage} &
        \begin{minipage}[t]{0.49\textwidth}
        \centering
        \includegraphics[width=\textwidth]{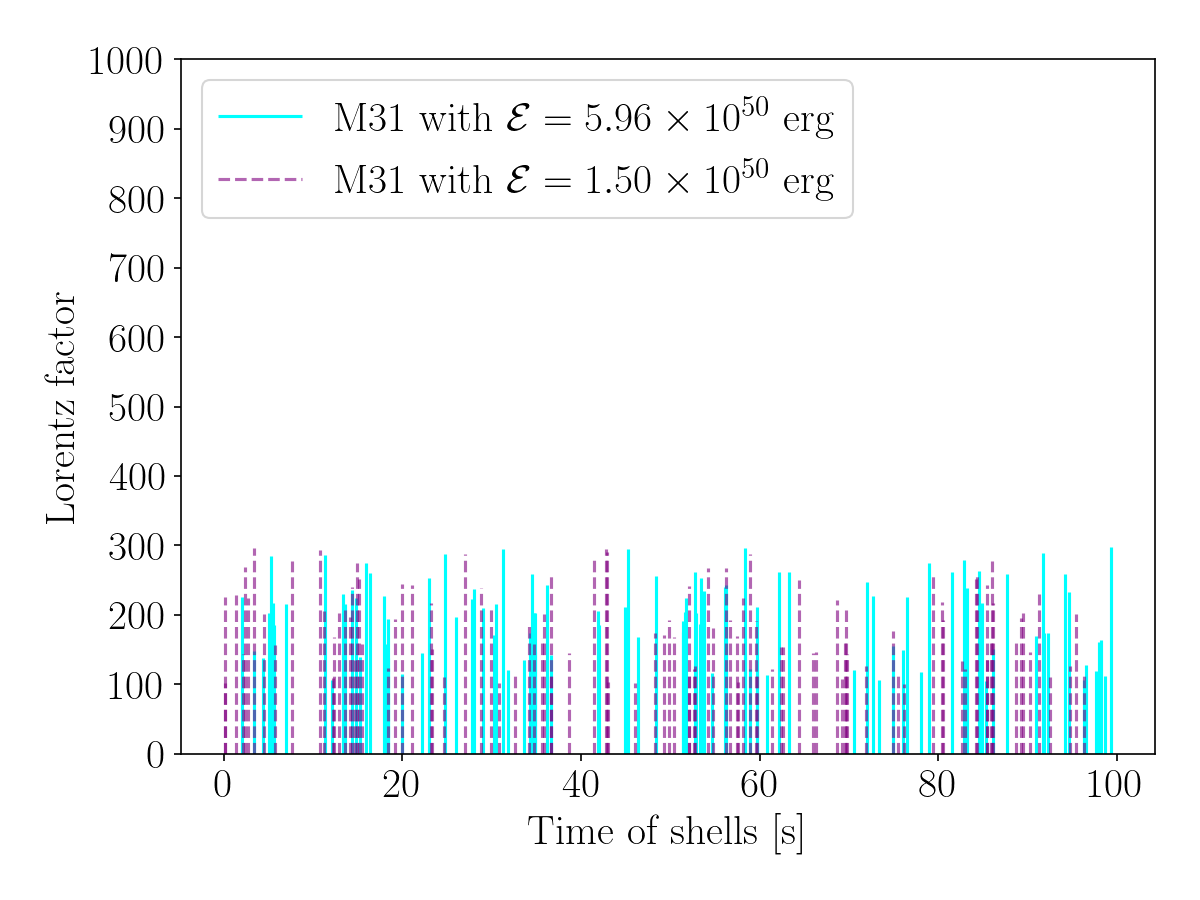}
      \end{minipage}
    \end{tabular}
    \caption{Hypothetical waveforms from M31. The source of distance is $R=0.77$~Mpc. The solid cyan line is $T_\mathrm{CE} = 97.3$~sec, and the dashed purple line is $T_\mathrm{CE} = 96.1$~sec. 
    All plots have  random $\gamma_j \in [100, 300]$, and $\delta \tau_j$ are applied for each shell.
    The other parameters are set as $r_0 = 10^7~\mathrm{cm}$, $N = 100$, and $W = 10^7~\mathrm{cm}$, $\Delta\theta = 0.05$ rad, and $\theta_v = 0.3$ rad.}
    \label{fig:m31}
\end{figure*}

We simulated GW memory from M31 and discussed its detectability. The distance to the source M31 is relatively close, $R=0.77$~Mpc, and we typically set the parameters such as $T_\mathrm{CE} \sim 100$~sec and $\gamma_j \in [100, 300]$. The hypothetical GW memory from M31 based on this event is shown in Fig.~\ref{fig:m31}.
Regarding the solid cyan line, we see that there is  the initial attenuation at $10^{-2}$~Hz, induced by $T_\mathrm{CE}^{-1} \sim 10^{-2}$~Hz, and further attenuation around $T_\mathrm{obs}^{-1} \sim 10^{3}$~Hz. The thickness impact appears around $\delta t_j^{-1} \sim 10^{3}$~Hz, although these amplitudes are under the sensitivity curve of ET-D.
The maximum amplitude given by Eq.~\eqref{eq:low-frew-approximately} is $\sim 10^{-23}$, which is almost the same as the signal amplitude in the frequency range observable by ET-D. This indicates that a GW memory observed by ET-D is helpful to understand the total energy $\mathcal{E}$ of GRBs, which is the same situation in Fig.~\ref{fig:GRB221009A}.
Furthermore, the detection by DECIGO and ET-D depends on the $\mathcal{E}$.
The dashed purple line in Fig.~\ref{fig:m31} shows the waveform under the same conditions as the solid cyan line, except for a smaller $\mathcal{E}$. 
This indicates that even for nearby GW events, detection remains challenging with next-generation detectors when the energy of the central engine is relatively small, around $10^{50}~\mathrm{erg}$.

\section{Summary and Discussion\label{sec:summary}}
\begin{figure}[htb]
    \centering
        \centering
        \includegraphics[width=\columnwidth]{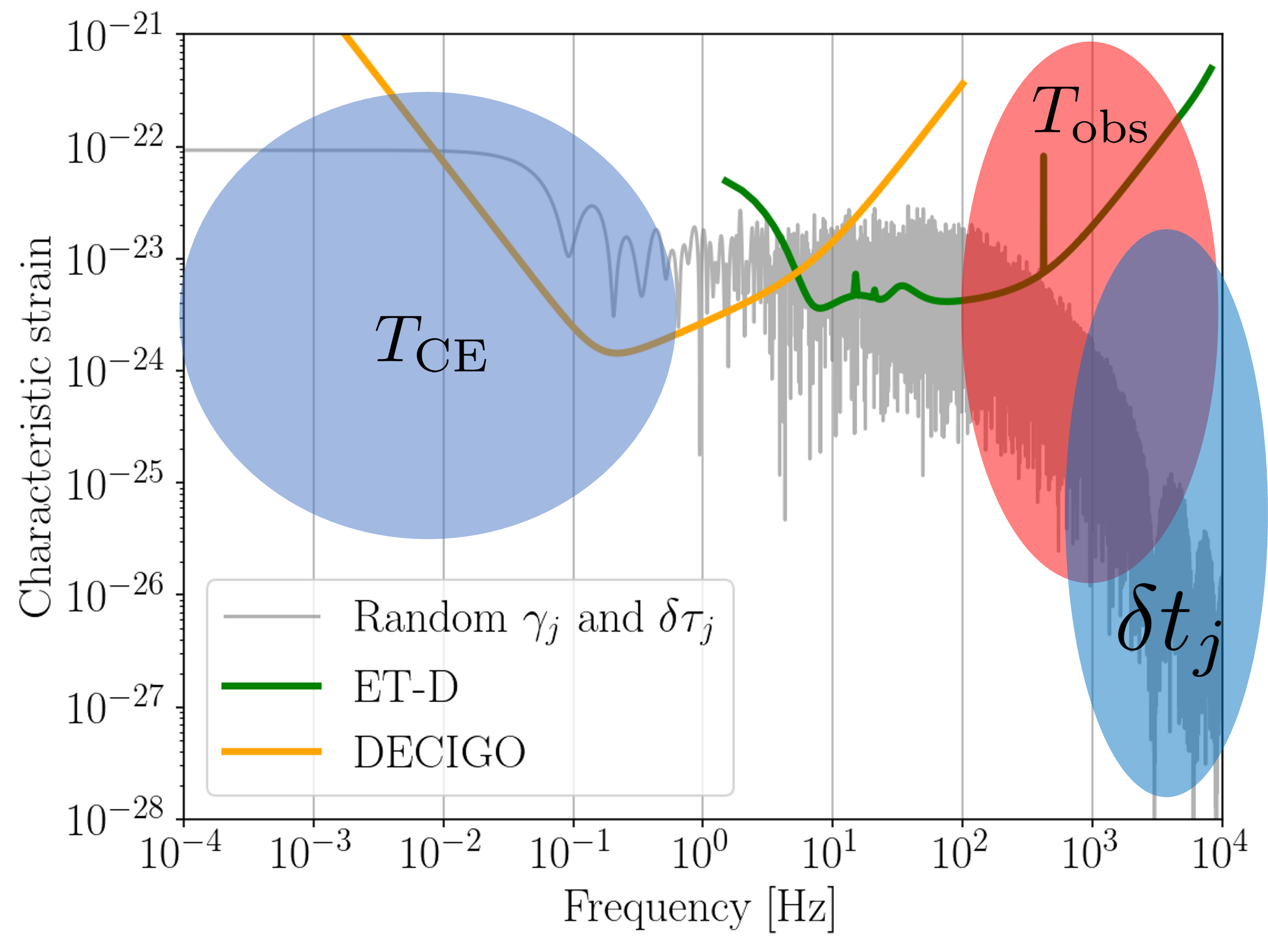}
    \caption{The illustration of relation between three physical times $T_{\mathrm{CE}}$, $\delta t_j$, and $T_{\mathrm{obs}}$ and waveform attenuation. The solid gray line is also plotted as a reference, which is the same waveform as in Fig.~\ref{fig:random-gamma-time-a}}
    \label{fig:relation-phys-attenuation}
\end{figure}

In this paper, we demonstrated the GW memory properties from multiple shells with several physical configurations. We have considered a thick shell with a radially uniform mass distribution. In the case $\theta_v\sim\Delta\theta$, the asymptotic form $f|h(f)|$ at the high-frequency end is proportional to $f^{-2}$.
In contrast, for a thin shell, such a relation at the high-frequency end follows $f|h(f)|\propto f^{-1}$. The asymptotic power-law index depends on the radial mass distribution \cite{ref:piran2021}. Hence, it might be possible to know the evolution of the mass ejection rate, although the frequency containing
the information is too high to be observed by ET-D.

The waveform produced by a single thick shell
with shell spreading 
is very similar to that produced without spreading. 
This indicates that the spreading has little effect on the waveform properties near the initial radius
$r_0\sim10^7~\mathrm{cm}$, that is the inner radius of the jet where the GW memory is emitted. In other words, the width of the spreading shell does not expand significantly in this region.

The time interval of each shell $\delta\tau_j$ with multiple shells affects the waveform shape. When the $\delta\tau_j = {\rm const}$, which means $W_j={\rm const}$, the waveform has multiple peaks at specific frequencies due to the vanishing attenuation. If such GW memories are observed, the shells must be emitted periodically. In contrast, the waveform loses such specific peaks when the $\delta\tau_j$ exhibits disorder, such as each shell emitted on a different interval time $\delta\tau_j$.

We also confirm that $f|\tilde{h}(f)|$ has a flat shape up to $f <$ ~kHz when the interval of multiple shells exhibits disorder. On the other hand, the waveform weakly depends on the terminal Lorentz factor, regardless of whether it is constant or exhibits disorder.

Several conditions involving $T_\mathrm{CE}$ and $\delta t_j$, and $T_\mathrm{obs}$, which represents a characteristic timescale as measured by an observer, have been examined. Figure~\ref{fig:relation-phys-attenuation} illustrates the relationship between these physical timescales and the waveform attenuation. We confirm that $T_\mathrm{CE}$ affects the waveform at low frequencies ($f \ll 1$~Hz), a range detectable by DECIGO. The parameter $\delta t_j$, which relates to the shell thickness, induces specific attenuation for frequencies above $\sim$~kHz. However, the attenuation due to the $T_\mathrm{obs}$ begins at lower frequencies than that due to $\delta t_j$, 
and this effect might be observable with ET-D.

Furthermore, we have shown that the GW memories from relatively energetic phenomena with typical GRB parameters would be detectable by multiple GW detectors, ET-D and DECIGO.\\

After the initial acceleration, the GW amplitude on the low-frequency side is saturated, which is determined by the total energy in the jet. In this sense, the GW memory can serve as a form of calorimetry.
However, after the initial acceleration, the jet evolves as it propagates as introduced in Sec.~\ref{sec:intro}.
The amplitude of the GW memory varies with time, and its detailed analysis may tell us the total jet energy.
Once the jet energy is known, we can estimate the efficiency of the prompt gamma-ray emission, whose constraints on the emission mechanism are currently unknown.
Note that the order of magnitude of the jet kinetic energy of low-luminosity GRBs has not yet been rigorously determined.
A popular model is in the low energy jet, while the off axis jet model gives larger energy \cite{ref:Yamazaki2003_offaxis,rerf:Ramirez-Ruiz2005,ref:Sato2021,ref:Sato2023}.



This paper shows that the GW memory from accelerating jets is only detectable for nearby sources.
However, multimessenger signals other than GW are likely to be expected, if the GW memory signatures are detected. In most cases, the jet may be viewed offaxis. Then, electromagnetic emission is dim in gamma rays, and typical internal shock emission may be seen in the x-ray or ultraviolet/optical band due to the relativistic beaming effect \cite{ref:Yamazaki2002,ref:Yamazaki2003CJ,ref:Yamazaki2018}. The prompt GRB emission may not be detected, and only off axis afterglow emission may be observable
\cite{ref:Nakar2002,ref:Totani2002,ref:Urata2015}.
If the GRB jet contains baryons to some extent, one can also expect high-energy (TeV--PeV) neutrinos \cite{ref:Waxman1997,ref:Guetta2004,ref:Murase2006,ref:Kimura2017,
ref:Kimura2022arXiv,
ref:Matsui2023}. When the jet is viewed off axis, the neutrino emission is again dim but may still be observable \cite{ref:Ahlers2019}.

\begin{acknowledgments}
We would like to thank Drs.~K.~Ioka, S.~J.~Tanaka, S.~Kimura and K.~Komori for their
valuable comments.
This research was supported in part by the Japan Society for the Promotion of Science (JSPS) Grant-in-Aid for Scientific Research [No. 23K22522 and No. 23H04899 (R.~Y.): No.  21H01082 and No. 23K20845 (N.~S.): No. 23H01176, 23K25872 and No. 23H04520 (H.~T.)] and Grant-in-Aid for JSPS Fellows [No. 22KF0329 (M.\ M.-C.)].
This research was supported by the Joint Research Program of the Institute for Cosmic Ray Research, University of Tokyo, and Tokyo City University Prioritized Studies.
\end{acknowledgments}

\section*{Data Availability}
The data that support the findings of this article are openly available~\cite{ref:data_availability}; embargo periods may apply.

\appendix

\section{NEXT GENERATION DETECTOR SENSITIVITIES}\label{app:next-gw-detectors}

\subsection{Einstein Telescope}
The ET~\cite{ref:Punturo2010} is an advanced third-generation GW detector proposed for construction underground in Europe. While its optical layout has not yet been finalized, various configurations have been proposed~\cite{ref:Branchesi2023}. The proposed configurations are based on an equilateral triangle arrangement of three detectors, each consisting of two interferometers with 10-km long arms. One of the two interferometers is optimized for low-frequency bands utilizing low laser power and cryogenic mirrors. In contrast, the other is optimized for high-frequency bands utilizing high laser power and mirrors at room temperature. We use the ``ET-D'' sensitivity curve~\cite{ref:Hild2011} which is provided by PyCBC software~\cite{ref:Pycbc}. We use this sensitivity curve divided by $\sin(60^\circ)$ to make it for a triangular interferometer because the original sensitivity is based on simulations for an L-shaped interferometer.

\subsection{DECIGO}
The Deci-hertz Interferometer Gravitational-Wave Observatory (DECIGO)~\cite{ref:Kawamura2008} is a space GW antenna, a Japanese challenging space mission. DECIGO will consist of four interferometer units in heliocentric orbit; each interferometer unit will consist of three drag-free spacecraft 1,000 km apart from each other. The DECIGO's target frequency band of GW is designed between 0.1 and 10 Hz, a challenging range for ground-based detectors to observe due to the seismic noise. Various innovative designs have recently been proposed to enhance its sensitivity~\cite{ref:galaxies11060111}. The DECIGO sensitivity curve that we use is provided by Eq.~(5) in Ref.~\cite{ref:decigo}.

\section{DERIVATION OF FORMULAS}\label{app:gwm}

\subsection{Derivation of Eq.~\eqref{eq:tacc}}

Let $r$ be the radial distance from the central engine, $r_0$ be the position where the jet begins to accelerate, and $\gamma$ be the Lorentz factor. The velocity divided by the speed of light, denoted as $\beta$, is used to express the Lorentz factor in terms of $\beta$ as follows:
\begin{equation}
\gamma = \cfrac{1}{\sqrt{1 - \beta^2}}
       = \cfrac{1}{\sqrt{1 - \left(\cfrac{1}{c}\cfrac{\mathrm{d}r}{\mathrm{d}t}\right)^2}}. \label{eq:lorentz-beta}
\end{equation}
Considering the situation of the accelerating jet, the Lorentz factor has a relation as follows:
\begin{equation}
\gamma = \frac{r}{r_0}, \label{eq:lorentz-position}
\end{equation} and $r = \gamma r_0$ denotes the position where the jet acceleration ends. We obtain an ordinary differential equation by Eq.~\eqref{eq:lorentz-beta} and Eq.~\eqref{eq:lorentz-position} as follows:
\begin{equation}
\frac{1}{\sqrt{1-\left(\cfrac{1}{c}\cfrac{\mathrm{d}r}{\mathrm{d}t}\right)^2}}
 = \frac{r}{r_0}.
\end{equation}
The acceleration time of the jet $t_\mathrm{acc}$ is written as
\begin{equation}
\int^{t_{\mathrm{acc}}}_0 \mathrm{d}t =\cfrac{1}{c}\int^{\gamma r_0}_{r_0} \cfrac{r\mathrm{d}r}{\sqrt{r^2- r_0^2}}.
\end{equation}
Then, we obtain
\begin{equation}
t_\mathrm{acc}(\gamma) = \frac{r_0}{c}\sqrt{\gamma^2-1}.
\end{equation}

\subsection{Derivation of Eq.~\eqref{eq:tobs}}
\begin{figure}[h]
    \centering
        \centering
        \includegraphics[width=\columnwidth]{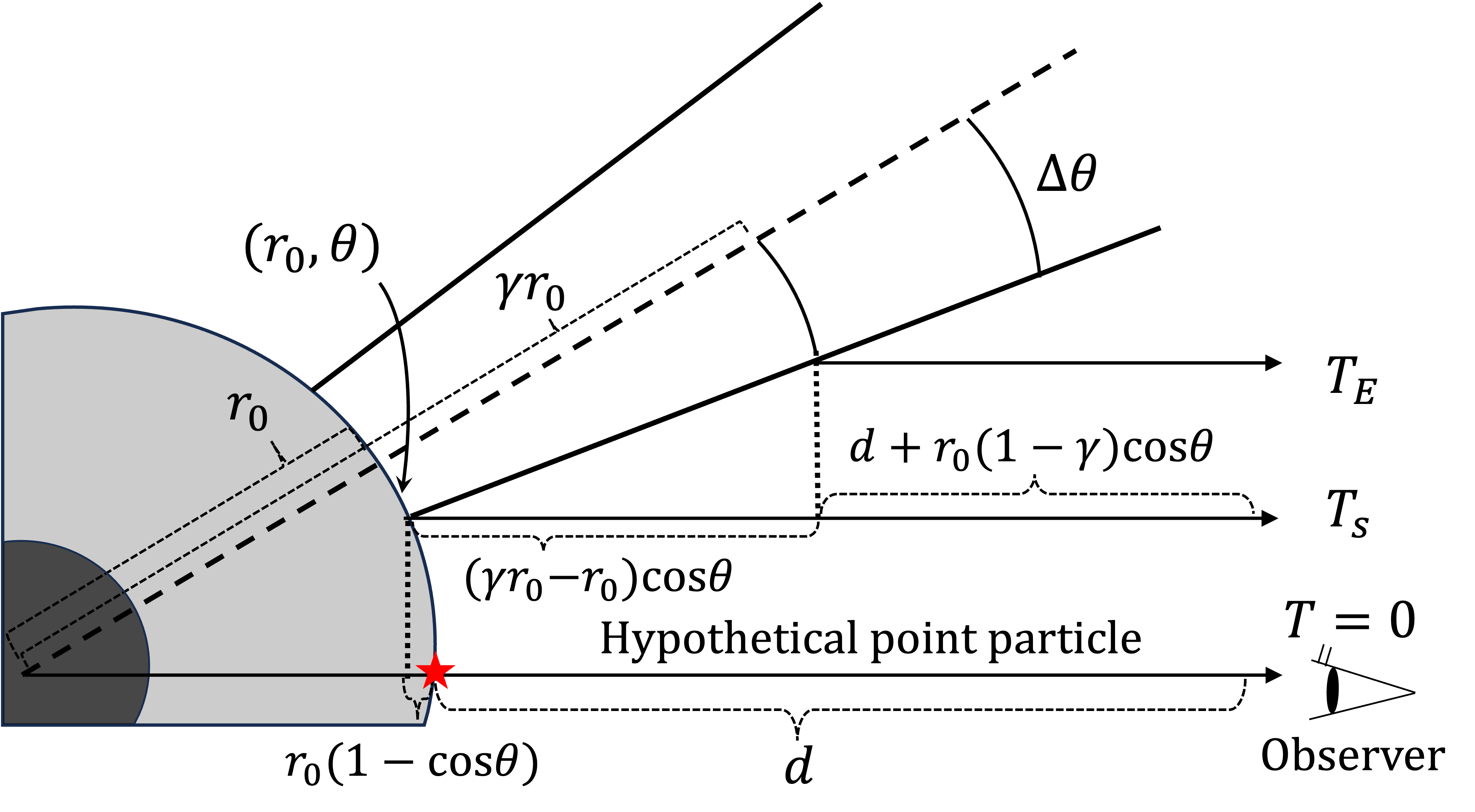}
    \caption{An illustration of an observer time.}
    \label{fig:observer-time}
\end{figure}

The geometric representation of the observer time is shown in Fig.~\ref{fig:observer-time}, where $d$ is the distance from the projected point to the observer, the point being along with an arc from the position of $r_0$ at angle $\theta$ to the line of a hypothetical point particle.
The observer start time $T_s(\theta)$ is expressed as
\begin{equation}
T_s(\theta) = \frac{r_0}{c}(1 - \cos\theta) + \frac{d}{c},
\end{equation}
and the observer end time $T_E(\theta, \gamma)$ is
\begin{equation}
T_E(\theta, \gamma) = t_{\mathrm{acc}}(\gamma) + \frac{r_0( 1 - \gamma \cos\theta)}{c} + \frac{d}{c}.
\end{equation}
The observer time $T_\mathrm{obs}(\theta, \gamma)$ is obtained by
\begin{equation}
T_\mathrm{obs}(\theta, \gamma)= T_E(\theta, \gamma) - T_s(\theta).
\end{equation}
Then, we obtain
\begin{equation}
    T_\mathrm{obs}(\theta, \gamma) = \frac{r_0}{c}\left(
    \sqrt{\gamma^2 - 1} - (\gamma - 1) \cos\theta
    \right).
\end{equation}

\subsection{Derivation of Eq.~\eqref{eq:gw-thick}}
The time interval between the front and rear ends of the thick shell is denoted as $\delta t$. Considering the difference between the rear shell, which has finished accelerating at $\gamma r_0$, and the front shell to be the thickness $W$, we obtain
\begin{equation}
W = \beta c \delta t,
\end{equation}
and $\delta t \sim W / c$ is the expression under the ultrarelativistic speed.

Suppose that there are $n$ thin shells, where the thin shells are equally distributed, and the mass is rewritten as $m = n m^{\prime}$.
We take the limit $n \to \infty$ in Eq.~\eqref{eq:gw-thick-small-n} by keeping the energy $\gamma nm^{\prime}c^2$ as constant. For simplicity, only terms that depend on the number of shells $k$ are considered, and the expression is as follows:

\begin{widetext}
\begin{equation}
\gamma m' c^2\sum^{n}_{k=0}e^{-i2\pi f\delta t k / n} = n\gamma m' c^2 \cdot \frac{1}{n}\left(1+\sum^{n}_{k=1}e^{-i2\pi f\delta t k / n}\right)
= \gamma m c^2 \left(\frac{1}{n}+\frac{1}{n}\sum^{n}_{k=1}e^{-i2\pi f \delta t k / n}\right),
\end{equation}
where we use $m = nm'$. We take the limit of $n\rightarrow\infty $, applying the piecewise quadrature method,
\begin{equation}
\lim_{n\rightarrow\infty} \gamma m' c^2\sum^{n}_{k=0}e^{-i2\pi f \delta t k / n} = \gamma m c^2 \left(\lim_{n\rightarrow\infty} \frac{1}{n}\sum^{n}_{k=1}e^{-i2\pi f \delta t k / n}\right)
= \gamma m c^2 \int^1_0 e^{-i2\pi f \delta t \cdot x} \ \mathrm{d}x
= i \gamma m c^2 \frac{e^{-i2\pi f\delta t}-1}{2\pi f \delta t}.
\label{eq:piecewise-quadrature}
\end{equation}
\end{widetext}


\bibliography{references}
\end{document}